\documentclass[
aip,
 amsmath,amssymb,
 reprint,
]{revtex4-1}
\usepackage{graphicx}% Include figure files
\usepackage{dcolumn}% Align table columns on decimal point
\usepackage{bm}% bold math
\usepackage{lipsum}
\usepackage[mathlines]{lineno}% Enable numbering of text and display math
\usepackage[utf8]{inputenc}
\usepackage{hyperref} % TODO please turn off when submitting
\usepackage[T1]{fontenc}
\usepackage{mathptmx}
\usepackage{etoolbox}
\makeatletter
\def\@email#1#2{%
 \endgroup
 \patchcmd{\titleblock@produce}
  {\frontmatter@RRAPformat}
  {\frontmatter@RRAPformat{\produce@RRAP{*#1\href{mailto:#2}{#2}}}\frontmatter@RRAPformat}
  {}{}
}%
\makeatother
\begin{document}
\title[Molecular free energies, rates, and mechanisms]{Molecular free energies, rates, and mechanisms from data-efficient path sampling simulations}
\author{Gianmarco Lazzeri}
\affiliation{Frankfurt Institute for Advanced Studies, Frankfurt am Main, Germany.}
\affiliation{Goethe University Frankfurt, Frankfurt am Main, Germany.}
\author{Hendrik Jung}
\affiliation{Goethe University Frankfurt, Frankfurt am Main, Germany.}
\affiliation{Department of Theoretical Biophysics, Max Planck Institute of Biophysics, Frankfurt am Main, Germany.}
\author{Peter G. Bolhuis}
\affiliation{Van 't Hoff Institute for Molecular Sciences, University of Amsterdam, Amsterdam, The Netherlands.}
\author{Roberto Covino*}
\affiliation{Goethe University Frankfurt, Frankfurt am Main, Germany.}
\affiliation{Frankfurt Institute for Advanced Studies, 60438 Frankfurt am Main, Germany.}
\email[Author to whom any correspondence should be addressed: ]{covino@fias.uni-frankfurt.de}
\date{\today}

\begin{abstract}
Molecular dynamics is a powerful tool for studying the thermodynamics and kinetics of complex molecular events. However, these simulations can rarely sample the required time scales in practice. Transition path sampling overcomes this limitation by collecting unbiased trajectories capturing the relevant events. Moreover, the integration of machine learning can boost the sampling while simultaneously learning a quantitative representation of the mechanism. Still, the resulting trajectories are by construction non-Boltzmann-distributed, preventing the calculation of free energies and rates. We developed an algorithm to approximate the equilibrium path ensemble from machine learning-guided path sampling data. At the same time, our algorithm provides efficient sampling, the mechanism, free energy, and rates of rare molecular events at a very moderate computational cost. We tested the method on the folding of the mini-protein chignolin. Our algorithm is straightforward and data-efficient, opening the door to applications on many challenging molecular systems.   
\end{abstract}

\maketitle

\begin{figure*}
\centering
\hspace{-4.5pt}
\includegraphics[width=.8\textwidth]{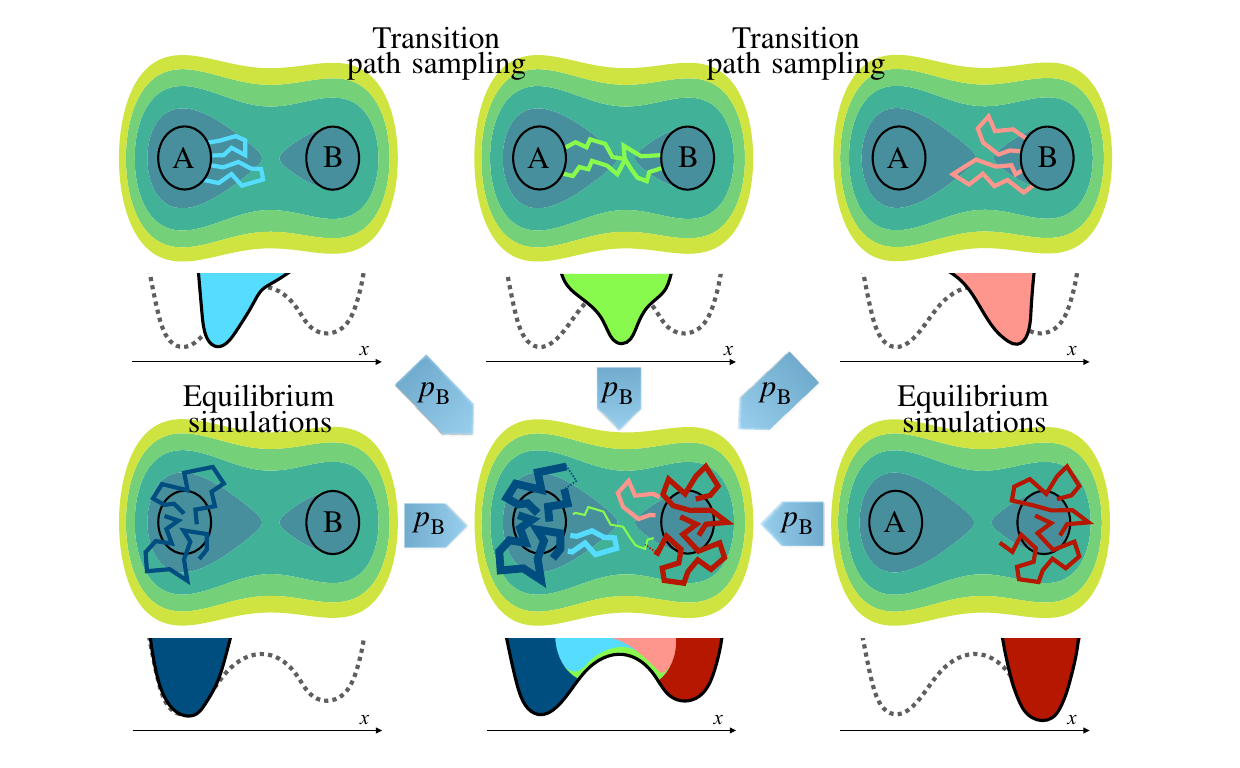}
\caption{Illustration of the algorithm. We simulate transition paths connecting two metastable states (top center) with AIMMD, which also generates excursions (top left and right) in the process. Meanwhile, we perform equilibrium simulations around the states (bottom left and right). Configurations in the trajectories generated by path sampling do not follow the Boltzmann distribution (dashed profile) by construction. After reweighting them based on a machine-learned committor $p_{\mathrm{B}}(x)$, we can merge all trajectories and recover the equilibrium kinetics and thermodynamics of the entire configuration space (bottom center).}
\label{fig:1}
\end{figure*}

\section{\label{sec:introduction}Introduction}

Molecules are everywhere. They constitute biological structures, chemical reactions, and materials. Molecules are also inherently dynamical.
Molecular dynamics (MD) simulations are accurate physics-based models that give access to the time evolution of molecular systems with atomic resolution, including complex biomolecules, materials, and chemical reactions \cite{dror2012biomolecular,massobrio2015molecular,frenkel2001understanding}. MD can provide a thermodynamic, kinetic, and mechanistic characterization of a wide range of phenomena, such as conformational changes\cite{adcock2006molecular,shaw2010atomiclevel}, folding\cite{swope2004describing,best2012atomistic}, ligand binding\cite{deng2009computations}, oligomerization\cite{psachoulia2010molecular,urbanc2010elucidation}, protein-membrane interactions\cite{nishimura_unique_2023}, nucleation\cite{matsumoto2002molecular}, and ion permeation\cite{allen2006molecular}.

Ideally, long MD simulations would produce equilibrium trajectories extensively exploring the configuration space of a molecular system \cite{wales2005energy,onuchic2004theory}. In these simulations, the trajectories would enter metastable states and spend most time there. Rarely, they would go on brief excursions in the transition region outside metastable states. Even more rarely, those excursions would result in an actual transition, crossing an energy barrier to reach an alternative state\cite{freddolino2010challenges}.

Only by repeatedly transitioning between metastable states these long trajectories would sample the stationary Boltzmann distribution that describes the system's thermodynamics\cite{peters2017reaction}. One could then count how often the trajectories undergo a transition to obtain the reaction rate constants that describe the kinetics. Moreover, one could isolate the trajectory segments that connect two metastable states---the transition paths (TPs)---and collect them in the transition path ensemble (TPE), which describes the mechanism of the transition\cite{vanden2010transition, roux_transition_2022}.

However, many interesting phenomena are rare events, stochastic transitions occurring on exponentially longer time scales than the MD integration time step\cite{hartmann2013characterization}. This makes them practically inaccessible by typical MD simulations\cite{lane2013milliseconds}. One solution is to apply an unphysical biasing force that steers the dynamics and enhances the exploration of the configuration space\cite{henin2022enhanced}. These methods require accurate prior knowledge of the system---a low-dimensional feature representation or, ideally, a reaction coordinate\cite{fiorin2013using,bernardi2015enhanced}.
Additionally, the bias distorts the system's dynamics, complicating its reconstruction
\cite{gershenson2020successes}.

Transition path sampling (TPS) and analogous methods \cite{zuckerman2017weighted} offer an alternative approach without adding any unphysical bias \cite{bolhuis2002transition}. TPS is a rigorous way to collect the TPs of a hypothetical extensive equilibrium trajectory \cite{vanden2010transition,metzner2009transition}. By avoiding sampling in the metastable states, TPS can be computationally very efficient\cite{bolhuis2002transition} and has enabled the characterization of several molecular processes \cite{bolhuis2003transition,knott2014mechanism,bolhuis2000reaction,vlugt2000diffusion,boulfelfel2015modeling,escobedo2009transition}. The resulting TPE contains mechanistic information\cite{weinan2005transition}.  In practice, generating TPs with high efficiency can be challenging, limiting the effectiveness of TPS schemes\cite{bolhuis2015practical}.  

Recently, we integrated deep learning with TPS to automatize and significantly improve the sampling of TPs in complex molecular systems \cite{jung2023machine}. We called this approach ``AI for molecular mechanism discovery'' (AIMMD). In AIMMD, a neural network controls TPS and boosts the production of TPs connecting two states; at the same time, it autonomously learns the transition mechanism by learning the committor\cite{jung2023machine}. The committor is the ideal reaction coordinate describing a general stochastic transition between two states, i.e., an optimal one-dimensional projection that quantitatively monitors the progress along a transition\cite{berezhkovskii2022relations}. 
However, TPS and AIMMD do not directly provide free energy profiles and rates \cite{bolhuis2015practical}. Configurations in the TPE are by construction not Boltzmann-distributed. 

Let us focus on a transition between two states, A and B, even though the following considerations are general. Thinking again at an ideal extensive trajectory, we can collect all trajectory segments that leave a state and end as soon as they reach a state\cite{rogal2010reweighted}. We can classify them according to their origin and destination: A-to-A, A-to-B, B-to-A, and B-to-B.
The TPE contains only those segments that connect the two states (A-to-B and B-to-A); it 
does not contain any excursions (A-to-A and B-to-B), those trajectory segments that temporarily leave a state and come back to it before reaching another one. However, these excursions significantly contribute to the Boltzmann distribution in the transition region between A and B. Regarding kinetics, the TPE gives access to the transition path time (the average duration of a TP) but not to the waiting times in the metastable states, which dominate the transition rate\cite{peters2017reaction}.

An effective strategy for obtaining the free energy is to use short, unbiased simulations. One solution is offered by transition interfaces sampling (TIS)\cite{van2005elaborating, hall_practical_2022}, a popular extension of TPS. Rogal et al. introduced the reweighted path ensemble (RPE), which reweights the individual TIS trajectories to approximate the free energy in the transition region\cite{rogal2010reweighted}. However, TIS is computationally relatively expensive and depends effectively on the knowledge of a reasonably good reaction coordinate. Recently, Brotzakis and Bolhuis proposed an algorithm to approximate the RPE by ``waste-recycling'' TPS simulations \cite{brotzakis2019approximating,frenkel2006wasterecycling}. 

In this study, inspired by concepts introduced in Refs.~\onlinecite{rogal2010reweighted,brotzakis2019approximating}, we propose a new  computational scheme 
that enables us to simultaneously access mechanisms, thermodynamics, and kinetics of stochastic rare event transitions. We show that building on the committor estimated by AIMMD, we can estimate free energy profiles and rates from just a few TPS simulations. We also developed a procedure for extending the estimate of the free energy in the transition region to the metastable states---hence to the whole accessible configuration space---with minimal additional computational cost (Fig.~\ref{fig:1}). We illustrated our method on two benchmark 2-dimensional systems with high energy barriers and multiple reactive channels and to the folding of the mini-protein chignolin\cite{harada2011exploring}. In all cases, we successfully determined the complete free energy profiles and rates in a small fraction of the computational resources required by a typical MD simulation.

The paper is organized as follows. In Sec.~\ref{sec:theory}, we provide a detailed explanation of the algorithm. In Sec.~\ref{sec:methods}, we introduce the studied systems and present the computational methods. In Sec.~\ref{sec:results}, we illustrate our results with particular emphasis on the computational performance. We end with concluding remarks and a future outlook.

\section{\label{sec:theory}Theory}

\subsection{\label{sec:theoryA} Summary of AIMMD}
For the sake of completeness, we briefly summarize the theory behind the AIMMD sampling scheme\cite{jung2023machine}. Let us consider a system with two metastable states $\mathrm{A}$ and $\mathrm{B}$ separated by an energy barrier\cite{peters2017reaction,vanden2010transition}. We assume that the system's dynamics (in the full configuration space) are Markovian. 
The system is described by its configuration $x$. 
A  trajectory, or path, is a sequence $\textbf{x} =\{x_0,~\dots,~x_t,~\dots,~x_L\}$
sampled at regular time intervals of length $L[\textbf{x}]$. 

TPS is a Markov chain Monte Carlo technique that generates a series of paths $\textbf y^{(1)},~\textbf y^{(2)},~\dots,~\textbf y^{(i)}$ connecting $\mathrm{A}$ and $\mathrm{B}$\cite{bolhuis2002transition,bolhuis2021transition}. In this work, $\textbf{x}^{(i)}$ is the trajectory simulated at step $i$, and $\textbf{y}^{(i)}$ the last accepted path after that step.
While the $\textbf{x}^{(i)}$ are always different, the $\textbf{y}^{(i)}$ can repeat in case of rejection.
Therefore, $\{\textbf y^{(1)},~\dots,~\textbf y^{(i)}\}$ is a subset of $\{\textbf x^{(1)},~\dots,~\textbf x^{(i)}\}$.
With an increasing number of steps, the chain of paths converges to the TPE (the equilibrium ensemble of all the system's TPs):
\begin{equation}
    \mathcal P_{\mathrm{TP}} \approx \left\{
    \textbf y^{(1)},~\textbf y^{(2)},~\dots,~\textbf y^{(n)}
    \right\}. 
\end{equation}

Element $\textbf y^{(i)}$ in the chain is generated from $\textbf y^{(i-1)}$. First, we select a shooting point $x^{(i)}_{\mathrm{sp}}$ from $\textbf y^{(i)}$ in the transition region. Then, we produce a trial path $\textbf x^{(i)}$ by a two-way shooting move\cite{mullen2015easy}: we evolve two sub-trajectories from $x^{(i)}_{\mathrm{sp}}$ backward and forward in time until they hit either $\mathrm{A}$ or $\mathrm{B}$, time-reverse the former, and join the two sub-trajectories together. To satisfy the fundamental requirement of detailed balance, the acceptance probability of $\textbf x^{(i)}$ as the next element in the chain is:
\begin{equation}
\begin{aligned}
    \label{acceptance}
    &p_{\mathrm{acc}}\left[\textbf y^{(i-1)} \rightarrow \textbf x^{(i)} \right] \\&\quad=
    \tilde h_{\mathrm{AB}}[\textbf x^{(i)}]~
    \min \left[ 1,
    \frac{p_{\mathrm{sel}}(x_{\mathrm{sp}}^{(i)};~ \textbf{x}^{(i)})}
         {p_{\mathrm{sel}}(x_{\mathrm{sp}}^{(i)}; ~\textbf{y}^{(i-1)})}\right].
\end{aligned}
\end{equation}
The indicator functional $\tilde h_{\mathrm{AB}}\left[\textbf{x}\right]$ 
equals unity 
if the trajectory connects $\mathrm{A}$ and $\mathrm{B}$, and zero  otherwise. $p_{\mathrm{sel}}(x_{\mathrm{sp}}^{(i)};~\textbf{x})$ is the probability of selecting $x^{(i)}_{\mathrm{sp}}$ among the configurations of trajectory $ \textbf{x}$; it can be any selection criterion function and can even change at different steps\cite{bolhuis2021transition}. The $\textbf y^{(i-1)}\rightarrow\textbf x^{(i)}\equiv\textbf y^{(i)}$ move is accepted or rejected according to $p_{\mathrm{acc}}$; in the latter case, we repeat $\textbf y^{(i)} \equiv \textbf y^{(i-1)}$. Note that the trial TPs may have $p_{\mathrm{acc}}$ lower than one and therefore get rejected, modifying the weights of the accepted trajectories. A good TPS algorithm increases the acceptance probability of the trial paths while preserving their 
heterogeneity\cite{bolhuis2015practical,falkner2022conditioning}.

In AIMMD, a neural network adaptively controls TPS in a data-driven way\cite{jung2023machine}. The network models the committor $p_{\mathrm{B}}(x)$---the probability that a trajectory initiated with random velocities at $x$ reaches $\mathrm{B}$ before $\mathrm{A}$\cite{berezhkovskii2019committors,peters2017reaction, roux_transition_2022}. The committor quantifies the progress along the transition and is considered the optimal reaction coordinate \cite{berezhkovskii2019committors, chen_discovering_2023}. This enables us to quantify the transition mechanism, and it also allows us to control the sampling. In fact, in the limit of Markovian dynamics, the probability of sampling a TP by a two-way shooting from $x$ is\cite{hummer2004transition}
\begin{equation}\label{pTP}
    P(\mathrm{TP}\mid x) = 2~p_{\mathrm{B}}(x)~(1-p_{\mathrm{B}}(x)).
\end{equation}
Since we do not apply bias forces to accelerate the transition in any direction, the dynamics remain time-reversible, such that we can always exchange $\mathrm{A}$ and $\mathrm{B}$, $p_\mathrm{B}$ and $p_\mathrm{A}=1-p_\mathrm{B}$.

In AIMMD, we control sampling by modeling the shooting point selection probability $p_{\mathrm{sel}}(x_{\mathrm{sp}}^{(i)};~\textbf{x})$ as a function of the committor. This choice enables us to control the exploitation-exploration dilemma. A selection probability peaked around the transition state, $p_{\mathrm{B}}=0.5$, would result in a high TPs generating efficiency (exploitation). On the other hand, discovering new reaction channels---new transition mechanisms---often requires selecting points close to the state boundaries (exploration). Here, we strike a balance between the two requirements by selecting shooting points following a uniform distribution as a function of $p_{\mathrm{B}}$. Consequently, at convergence, the optimal sampling rate of new TPs will be $\langle P(\mathrm{TP}\mid x_{\mathrm{sp}}) \rangle_{p_{\mathrm{B}}}~=~1/3$.

We learn the committor by training a neural network. At each TPS step, we compare the expected and actual outcomes of the sub-trajectories originating from the shooting points.
Each shooting point $x_{\mathrm{sp}}^{(i)}$ returns $r^{(i)}$, an integer between 0 and 2 specifying how many sub-trajectories reached $\mathrm{B}$ before $\mathrm{A}$. The trained model maximises the likelihood of the $(x_{\mathrm{sp}}^{(i)},~r^{(i)})$ outcomes by minimizing the binomial loss\cite{peters2010recent,jung2023machine}
\begin{equation}\label{loss}
\begin{aligned}
    L^{(n_{\mathrm{TPS}})} &= - \sum_{i=1}^{n_{\mathrm{TPS}}}~v^{(i)}~\Bigg[r^{(i)}~\log p_{\mathrm{B}}(x_{\mathrm{sp}}^{(i)}) \\  & +~(2-r^{(i)})~\log (1-p_{\mathrm{B}}(x_{\mathrm{sp}}^{(i)}))\Bigg],
\end{aligned}
\end{equation}
where $n_{\mathrm{TPS}}$ is the number of trial shots, and $v^{(i)}$ is the importance of the $i$-th point (for the choice of $v^{(i)}$, see Section~\ref{sec:methodsA}).
In this way, the network learns the committor with no prior information and simultaneously enhances TPS.

\subsection{\label{sec:theoryB}Approximating the equilibrium path ensemble}

\begin{figure}
\centering
\hspace{-4.5pt}
\includegraphics[width=\columnwidth]{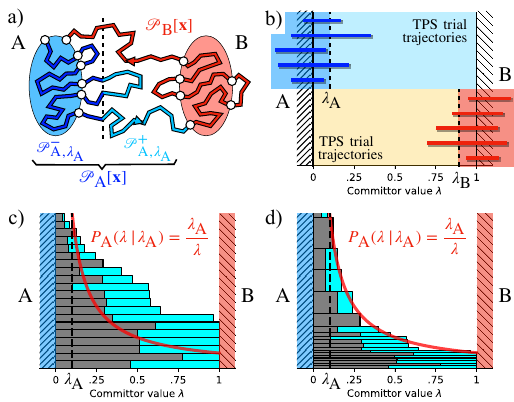}
\caption{\textbf{a)} Ensemble definitions. Given an infinitely long unbiased simulation, the path ensemble (PE) collects all the trajectory segments starting and ending upon crossing the boundary of a state (cuts at the white dots). The trajectories in $\mathcal P_{\mathrm{A}}$ (blue) start at the boundary of $\mathrm{A}$; their configurations populate $\mathcal P_{\mathrm{A},\lambda_{\mathrm{A}}}^+$ (light blue) and $\mathcal P_{\mathrm{A},\lambda_{\mathrm{A}}}^-$ (dark blue), based on whether they have committor values bigger or smaller than $\lambda_{\mathrm{A}}$. $\mathcal P_{\mathrm{B}}$ and $\mathcal P_{\mathrm{B},\lambda_{\mathrm{B}}}^\pm$ are defined analogously. \textbf{b)} Extension to the metastable states. The weighted TPS trajectories approximate the PE in the cyan ($\mathcal P_{\mathrm{A},\lambda_{\mathrm{A}}}^+$) and yellow ($\mathcal P_{\mathrm{B},\lambda_{\mathrm{B}}}^-$) regions. In the blue ($\mathcal P_{\mathrm{A},\lambda_{\mathrm{A}}}^-$) and red ($\mathcal P_{\mathrm{B},\lambda_{\mathrm{B}}}^+$) regions, we use (potentially) short simulations initiated around the states (horizontal lines); their occasional excursions in the TPS regions are crucial to the extension, as they determine the trajectories' relative weight. \textbf{c)} AIMMD trial trajectories to/from A before weighting. Each path is represented by a rectangle between its minimum and maximum committor values; the shooting point lies at the interface between the gray and cyan area. The trajectories' ``crossing statistics'' do not match the expected crossing probability for $\mathcal P_{\mathrm{A},\lambda_{\mathrm{A}}}^+$ (red line). \textbf{d)} The same trajectories as in (c) after applying the weighting scheme of Eq.~\eqref{weightsA}. The heights of the rectangles correspond to the weights of the paths. The crossing statistics now matches $P_{\mathrm{A}}(\lambda|\lambda_{\mathrm{A}})=\lambda_{\mathrm{A}}/\lambda$. }
\label{fig:2}
\end{figure}

AIMMD directly provides a valid estimate of the TPE. However, the TPs are only a small subset of the more general equilibrium path ensemble (PE) $\mathcal P[\textbf x]$, which  
consists of all unbiased trajectories that start and end as soon as they cross any state boundary. In addition to transitions and excursions, as defined in the Introduction, the PE also has trajectories entirely contained in either $\mathrm{A}$ or $\mathrm{B}$. We can split $\mathcal P[\textbf x]$ into $\mathcal P_{\mathrm{A}}[\textbf x]$ and $\mathcal P_{\mathrm{B}}[\textbf x]$---the path ensembles ``gravitating'' around the basins of attraction of states $\mathrm{A}$ and $\mathrm{B}$, respectively\cite{van2003novel}. $\mathcal P_{\mathrm{A}}$ (or $\mathcal P_{\mathrm{B}}$) contains all trajectories that start either entering or leaving A (or B) and end upon crossing any state boundary (A or B). Trajectories in $\mathcal P_{\mathrm{A}}[\textbf x]$ do not have configurations in $\mathrm{B}$ and vice-versa.

Under the ergodic hypothesis, sampling $\mathcal P[\textbf x] = \mathcal P_{\mathrm{A}}[\textbf x] \cup \mathcal P_{\mathrm{B}}[\textbf x]$ is equivalent to simulating and then splitting an infinitely long unbiased trajectory (Fig.~\ref{fig:2}a). Thus, the PE contains both thermodynamic and kinetic information about the studied transition.

Our goal is to approximate the PE with  a set of $n$ dynamically unbiased short trajectory segments, which are not necessarily a priori distributed  according to the equilibrium PE, and need to be reweighted. For this purpose, we use the following notation:
\begin{equation}\label{eq:abstract_ensemble}
    \mathcal P \approx \{(\textbf w^{(1)},~\textbf x^{(1)}),~\dots,~(\textbf w^{(i)},~\textbf x^{(i)}),~\dots,~(\textbf w^{(n)},~\textbf x^{(n)})\},
\end{equation}
where $\textbf w^{(i)}$ is the vector of weights associated with each configuration in trajectory $\textbf x^{(i)}$. 
%We do not require that all $\textbf x^{(i)}$ are proper elements of the PE. 
The aim of this approximation is that the distribution of the configurations in the reweighted trajectories, $\rho(x)$, must follow the Boltzmann distribution of the configurations in the PE. More generally, the ensemble average of any thermodynamic observable $O(x)$ must be
\begin{equation}\label{eq:abstract_projection}
    \langle O\rangle \approx \sum_{i=1}^n \sum_{t=0}^{L[\textbf x^{(i)}]} w^{(i)}_t~O(x^{(i)}_t).
\end{equation}

Given a (small) committor threshold $\lambda_{\mathrm{A}} > 0$, we further split $\mathcal P_{\mathrm{A}}[\textbf x]$ into $\mathcal P_{\mathrm{A},\lambda_{\mathrm{A}}}^-$ and $\mathcal P_{\mathrm{A},\lambda_{\mathrm{A}}}^+$. The configurations in  
$\mathcal P_{\mathrm{A},\lambda_{\mathrm{A}}}^-$ have committor values smaller than $\lambda_{\mathrm{A}}$, while those in $\mathcal P_{\mathrm{A},\lambda_{\mathrm{A}}}^+$ have $p_B(x) \geq \lambda_{\mathrm{A}}$ (Fig.~\ref{fig:2}a). Given a (large) committor threshold $\lambda_{\mathrm{B}} < 1$, we obtain $\mathcal P_{\mathrm{B},\lambda_{\mathrm{B}}}^-$ and $\mathcal P_{\mathrm{B},\lambda_{\mathrm{B}}}^+$ analogously. These new ensembles are strictly speaking not proper path ensembles but are proper configurational ones.

$\mathcal P_{\mathrm{A},\lambda_{\mathrm{A}}}^-$, $\mathcal P_{\mathrm{A},\lambda_{\mathrm{A}}}^+$, $\mathcal P_{\mathrm{B},\lambda_{\mathrm{B}}}^-$, and $\mathcal P_{\mathrm{B},\lambda_{\mathrm{B}}}^+$ form a partition of configurations in the PE. We will approximate each one separately and join them together once determined their relative weights. 

The configurations in $\mathcal P_{\mathrm{A},\lambda_{\mathrm{A}}}^+$, $\mathcal P_{\mathrm{B},\lambda_{\mathrm{B}}}^-$  are more difficult to sample, since they contain the rare event of interest. Our approach is to approximate the $\mathcal P_{\mathrm{A},\lambda_{\mathrm{A}}}^+$, $\mathcal P_{\mathrm{B},\lambda_{\mathrm{B}}}^-$ ensembles with the $n_{\mathrm{TPS}}$ trial paths $\textbf x^{(1)},~\dots,~\textbf x^{(n_{\mathrm{TPS}})}$ produced by AIMMD to sample the transition between A and B. Crucially, these include the paths that
were not reactive ($\mathrm{A}$-to-$\mathrm{A}$ and $\mathrm{B}$-to-$\mathrm{B}$). The justification for this lies in the path-recycling method introduced in Ref.~\cite{brotzakis2019approximating}, which established that trial trajectories created by two-way shooting, whether transitions or not, are
proper paths that take part in the equilibrium PE. By creating the trial paths along the entire order parameter range, one ensures proper coverage of the PE.
The resulting trial paths are naturally not distributed according to the equilibrium ensemble because they were created from a biased selection and thus must be properly reweighted. 

For $\mathcal P_{\mathrm{A},\lambda_{\mathrm{A}}}^-$ and $\mathcal P_{\mathrm{B},\lambda_{\mathrm{B}}}^+$, which entirely include the metastable states, we will use $n_{\mathrm{A}}+n_{\mathrm{B}}$ short unbiased trajectories $\textbf z_{\mathrm{A}}^{(1)},~\dots,~\textbf z_{\mathrm{A}}^{(n_{\mathrm{A}})}$ and $\textbf z_{\mathrm{B}}^{(1)},~\dots,~\textbf z_{\mathrm{B}}^{(n_{\mathrm{B}})}$ initialized around $\mathrm{A}$ and $\mathrm{B}$, respectively (Fig.~\ref{fig:2}b). In this way, we complement the TPS trajectories with short equilibrium simulations and extend $\mathcal P$ to the metastable states. 

Our estimate of the properly weighted configurations in the PE becomes
\begin{equation}\label{ensemble}
\begin{aligned}
    \mathcal P \approx~&\{(\textbf v_{\mathrm{A}}^{(1)},~\textbf z_{\mathrm{A}}^{(1)}),~\dots,~(\textbf v_{\mathrm{A}}^{(n_{\mathrm{A}})},~\textbf z_{\mathrm{A}}^{(n_{\mathrm{A}})})\}&\cup \\
    & \{(\textbf w_{\mathrm{A}}^{(1)},~\textbf x^{(1)}),~\dots,~(\textbf w_{\mathrm{A}}^{(n_{\mathrm{TPS}})},~\textbf x^{(n_{\mathrm{TPS}})})\}&\cup \\
    & \{(\textbf w_{\mathrm{B}}^{(1)},~\textbf x^{(1)}),~\dots,~(\textbf w_{\mathrm{B}}^{(n_{\mathrm{TPS}})},~\textbf x^{(n_{\mathrm{TPS}})})\}&\cup \\
    &\{(\textbf v_{\mathrm{B}}^{(1)},~\textbf z_{\mathrm{B}}^{(1)}),~\dots,~(\textbf v_{\mathrm{B}}^{(n_{\mathrm{B}})},~\textbf z_{\mathrm{B}}^{(n_{\mathrm{B}})})\},
\end{aligned}
\end{equation}
with $n=n_{\mathrm{TPS}} + n_{\mathrm{A}} + n_{\mathrm{B}}$. The first two sets jointly approximate $\mathcal P_{\mathrm{A}}$, and the latter two $\mathcal P_{\mathrm{B}}$, with the corresponding distributions of configurations $\rho_{\mathrm{A}}$ and $\rho_{\mathrm{B}}$.

\subsection{\label{sec:theoryC}Reweighting the TPS trial trajectories}

In this section, we derive a solution for the $\textbf w_{\mathrm{A}}^{(i)}$ and $\textbf w_{\mathrm{B}}^{(i)}$ vectors. We simplify the problem by assuming that all configurations in each trajectory are
weighted by the same factor
% uniformly weighted
within $\mathcal P_{\mathrm{A},\lambda_{\mathrm{A}}}^+$ and $\mathcal P_{\mathrm{B},\lambda_{\mathrm{B}}}^-$, i.e.:
\begin{subequations}
\begin{align}
    w_{\mathrm{A},~t}^{(i)} &= w_{\mathrm{A}}^{(i)}~\theta(p_{\mathrm{B}}(x^{(i)}(t)) \geq \lambda_{\mathrm{A}}),\\
    w_{\mathrm{B},~t}^{(i)} &= w_{\mathrm{B}}^{(i)}~\theta(p_{\mathrm{B}}(x^{(i)}(t)) \leq \lambda_{\mathrm{B}}),\quad\forall i,t
\end{align}
\end{subequations}
where $\theta(x)$ is the Heaviside function, which is $1$ if $x>0$ and $0$ otherwise.  Furthermore, $w_{\mathrm{A}}^{(i)}=0$ if $\textbf x^{(i)}$ does not originate and/or terminate in $\mathrm{A}$, and $w_{\mathrm{B}}^{(i)}=0$ if it does not start and/or end in $\mathrm{B}$.

Each TP has both $w_{\mathrm{A}}^{(i)} > 0$ and $w_{\mathrm{B}}^{(i)} > 0$. It contributes to both $\mathcal P_{\mathrm{A},\lambda_{\mathrm{A}}}^+$ and $\mathcal P_{\mathrm{B},\lambda_{\mathrm{B}}}^-$ because of microscopic time reversibility: an unbiased trajectory that goes from $\mathrm{A}$ to $\mathrm{B}$ is equivalent to the time-reversed counterpart from $\mathrm{B}$ to $\mathrm{A}$. By allowing for time-reversed trajectory segments, we improve the accuracy of the $\mathcal P_{\mathrm{A},\lambda_{\mathrm{A}}}^+$ and $\mathcal P_{\mathrm{B},\lambda_{\mathrm{B}}}^-$ estimates. Moreover, the weights will be halved to avoid double counting.

We introduce the
crossing probability
$P_{\mathrm{A}}(\lambda \mid \lambda_{\mathrm{A}})$\cite{van2003novel,van2005elaborating,cabriolu2017foundations}: the probability that a trajectory starting in $\mathrm{A}$ and crossing $\lambda_{\mathrm{A}} \in (0, \lambda]$ reaches $\lambda$, before returning to $\mathrm{A}$ or ending in $\mathrm{B}$.
While usually $\lambda$ is interpreted as an order parameter that is a reasonable proxy for the reaction coordinate, here we take $\lambda$  to be the best possible reaction coordinate, the committor itself.
By using  the committor we obtain a simple closed solution for the crossing probability (see  Appendix \ref{app1} for a proof):
\begin{equation}
\label{crossing0}
    P_{\mathrm{A}}(\lambda\mid\lambda_{\mathrm{A}}) = \frac{\lambda_{\mathrm{A}}}{\lambda},\quad \forall \lambda_{\mathrm{A}}\in (0,\lambda],~\forall \lambda \in [\lambda_{\mathrm{A}}, 1]
\end{equation}
The trajectory segments in $\mathcal P_{\mathrm{A},\lambda_{\mathrm{A}}}^+$ start and/or end in $\mathrm{A}$ and cross $\lambda_{\mathrm{A}}\ll 1$ by definition. Therefore, the fraction of paths reaching $\lambda$ must correspond to Eq.~\eqref{crossing0}. In particular, most trajectories would only make small excursions in the transition region ($\lambda\ll 1$) before returning to $\mathrm{A}$. We will approximate $\mathcal P_{\mathrm{A},\lambda_{\mathrm{A}}}^+$ with a finite set of trajectories. Again, the fraction of paths reaching $\lambda$ (the ``crossing statistics'') should match $P_{\mathrm{A}}(\lambda\mid \lambda_{\mathrm{A}})$.

In AIMMD, however, we initialize trajectories at higher committor values by controlling the selection probability of the shooting points. For example, let $\lambda^{(i)} \equiv p_{\mathrm{B}}(x_{\mathrm{sp}}^{(i)})$ be the value of the $i$-th shooting point. If $x_{\mathrm{sp}}^{(i)}$ is at the transition state ($\lambda^{(i)}=0.5$), then $\textbf x^{(i)}$ would start by construction at $\lambda=0.5$. This selection biases the crossing statistics (Fig.~\ref{fig:2}c). Only by appropriately weighting the trajectories can we match the observed statistics with the expected one and reconstruct $\mathcal P_{\mathrm{A},\lambda_{\mathrm{A}}}^+$.

The weights $w^{(i)}_{\mathrm{A}}$ should increase the contribution of small excursions and decrease the contribution of large excursions and TPs (Fig.~\ref{fig:2}c). 
The RPE theory\cite{rogal2010reweighted} demonstrates that the weight of each path $\textbf x^{(i)}$ depends on the furthermost value of the committor $\lambda_{\max}^{(i)}$ along that path (the magnitude of the associated excursion).

Intuitively, the reweighting is an importance sampling procedure:
the weight of $\textbf x^{(i)}$
should be of the form $E(\lambda_{\max}^{(i)})/S(\lambda_{\max}^{(i)})$, where $E(\lambda_{\max}^{(i)})$ is the fraction of expected paths that should touch at least $\lambda_{\max}^{(i)}$, and $S(\lambda_{\max}^{(i)})$ is the fraction of simulated paths that touched at least $\lambda_{\max}^{(i)}$.
From Eq.~\eqref{crossing0}, we know that $E(\lambda_{\max}^{(i)})$ is proportional to $1/\lambda_{\max}^{(i)}$. In this way, smaller excursions in the transition region get increasingly promoted as $\lambda_{\max}^{(i)}$ approaches $\lambda_{\mathrm{A}}$, while TPs ($\lambda_{\max}=1$) have the lowest weights. 
Conversely, $S(\lambda_{\max}^{(i)})$ is proportional to $m_{\mathrm{A}}(\lambda_{\max}^{(i)})$: the number of paths from $\mathrm A$ that touched $\lambda_{\max}^{(i)}$. 
However, for assessing $S$ we can only consider the trajectories shot from committor values $p_{\mathrm{B}} (x_{\mathrm{sp}}^{(j)}) = \lambda^{(j)}$ lower than $ \lambda_{\max}^{(i)}$---the only ones following the crossing statistics at $\lambda_{\max}^{(i)}$, since paths with a shooting point beyond $\lambda_{\max}^{(i)}$ 
are pushed closer to state B by construction. For these paths, we compute
\begin{equation}\label{mA}
m_{\mathrm{A}}(\lambda_{\max}^{(i)})=\sum_{j=1}^{n_{\mathrm{TPS}}} \tilde h_{\mathrm{A}}[\textbf x^{(j)}]~\theta(\lambda_{\max}^{(i)} - \lambda^{(j)})~\theta(\lambda_{\max}^{(j)}-\lambda_{\max}^{(i)}),
\end{equation}
which counts the trajectories generated from shooting points with committor values smaller than $\lambda_{\max}^{(i)}$ and reaching (at least) $\lambda_{\max}^{(i)}$. The indicator functional $\tilde h_{\mathrm{A}}$ ensures that $\textbf x^{(i)}$ starts/ends in $\mathrm{A}$ and crosses $\lambda_{\mathrm{A}}$. 

By matching the expected and observed crossing statistics, we finally obtain an analytical form for the weights
\begin{subequations}\label{weights}
\begin{equation}
\label{weightsA}
    w^{(i)}_{\mathrm{A}} = \tilde h_{\mathrm{A}}[\textbf x^{(i)}]~\frac{c_{\mathrm{A}}}{\lambda_{\max}^{(i)}~m_{\mathrm{A}}(\lambda_{\max}^{(i)})},
\end{equation}
where $c_{\mathrm{A}}$ is a normalizing constant. A consequence of Eq.~\eqref{weightsA} is that TPs are all reweighted by the same amount. In Appendix \ref{app3}, we show that Eq.~\eqref{weightsA} can be obtained rigorously as a limit case of the RPE theory\cite{rogal2010reweighted}. Switching states $\mathrm{A}$ and $\mathrm{B}$, we must exchange $p_{\mathrm{B}}(x)=\lambda$ with $p_{\mathrm{A}}(x) = 1 - \lambda$, but the derivation remains the same.
Thus
\begin{equation}
\label{weightsB}
    w_{\mathrm{B}}^{(i)} = \tilde h_{\mathrm{B}}[\textbf x^{(i)}]~\frac{c_{\mathrm{B}}}{(1-\lambda_{\min}^{(i)})~m_{\mathrm{B}}(\lambda_{\min}^{(i)})},
\end{equation}
\end{subequations}
and analogously, $m_{\mathrm{B}}(\lambda)$ counts the trajectories shot at committor values bigger than $\lambda$, starting and/or ending in $\mathrm{B}$, and reaching $\lambda$, while $\lambda_{\min}^{(i)}$ is the furthermost committor value reached by $\textbf x^{(i)}$ from $\mathrm{B}$ ($\lambda_{\mathrm{min}}=0$ if the path is reactive). In Figs.~\ref{fig:2}c,d, we show how this procedure recovers the expected crossing probability on synthetic data. A uniform shooting point distribution in committor space homogenizes $m_{\mathrm{A}}$ and $m_{\mathrm{B}}$ in Eq.~\eqref{weights} and improves the accuracy of the estimate.

The relative importance of $\mathcal P_{\mathrm{A}}$ and $\mathcal P_{\mathrm{B}}$ is set by the normalizing constants $c_{\mathrm{A}}$ and $c_{\mathrm{B}}$. We impose the fixed ratio:
\begin{equation}\label{normalization}
    \frac{c_{\mathrm{A}}}{c_{\mathrm{B}}} = \frac{\rho_{\mathrm{B}}'(\lambda = 0.5)}{\rho_{\mathrm{A}}'(\lambda=0.5)},
\end{equation}
where $\rho_{\mathrm{A}}'(\lambda=0.5)$ and $\rho_{\mathrm{B}}'(0.5)$ are the unnormalized $\mathcal P_{\mathrm{A}}$ and $\mathcal P_{\mathrm{B}}$ densities at the transition state. (For a  justification of Eq.~\eqref{normalization}, see Appendix \ref{app2}). 
In practice, we can compute the weighted population of the ensembles between, e.g., $\lambda=0.45$ and $\lambda=0.55$.
Using these constants thus ensures that the transition state interface has equal amounts of trajectories going to $\mathrm{A}$ and $\mathrm{B}$ in the PE.

\subsection{\label{sec:theoryD}Extension to the metastable states}

So far, we determined the weights for the paths in the transition region. Now we will 
derive a solution for the $\textbf v_{\mathrm{A}}^{(i)}$ and $\textbf v_{\mathrm{B}}^{(i)}$ vectors containing the weights for the short, unbiased simulations in the wells. As the $\textbf z_{\mathrm{A}}^{(i)}$ trajectories are the outcome of equilibrium sampling around state $\mathrm{A}$, their configurations must have equal weight throughout $\mathcal P_{\mathrm{A}, \lambda_{\mathrm{A}}}^-$; the same argument holds for the $\textbf z_{\mathrm{B}}^{(i)}$. Thus:
\begin{subequations}
\begin{align}
    v_{\mathrm{A},~t}^{(i)} &= \gamma_{\mathrm{A}}~\theta(p_{\mathrm{B}}(z_{\mathrm{A}}^{(i)}(t)) < \lambda_{\mathrm{A}}),\\
    v_{\mathrm{B},~t}^{(i)} &= \gamma_{\mathrm{B}}~\theta(p_{\mathrm{B}}(z_{\mathrm{B}}^{(i)}(t)) > \lambda_{\mathrm{B}}),\quad\forall i,t
\end{align}
\end{subequations}
where $\gamma_{\mathrm{A}}$ and $\gamma_{\mathrm{B}}$ are positive constants.
The occasional excursions of the $\textbf z_{\mathrm{A}}^{(i)}$ and $\textbf z_{\mathrm{B}}^{(i)}$
above and below the $\lambda_{\mathrm{A}}$ and $\lambda_{\mathrm{B}}$ thresholds are instrumental for determining $\gamma_{\mathrm{A}}$ and $\gamma_{\mathrm{B}}$ and therefore extending our $\mathcal P[\textbf x]$ evaluation to the metastable states. The number of $\textbf z^{(i)}_{\mathrm{A}}$ configurations that go beyond $\lambda_{\mathrm{A}}$, when multiplied by $\gamma_{\mathrm{A}}$, must match the total population of $\mathcal P_{\mathrm{A}, \lambda_{\mathrm{A}}}^+$. Conversely, the number of $\textbf z^{(i)}_{\mathrm{B}}$ configurations that cross $\lambda_{\mathrm{B}}$, when multiplied by $\gamma_{\mathrm{B}}$, must correspond to $\mathcal P_{\mathrm{B}, \lambda_{\mathrm{B}}}^+$. We enforce the above statements by setting
\begin{subequations}\label{gamma}
    \begin{align}
        \gamma_{\mathrm{A}}(\lambda_{\mathrm{A}}) &=
        \frac{
            \sum_{i=1}^{n_{\mathrm{TPS}}}
            w_{\mathrm{A}}^{(i)}~
            \sum_{t=0}^{L[\textbf x^{(i)}]}
            \theta(p_{\mathrm{B}}(x^{(i)}(t)) \geq \lambda_{\mathrm{A}})
        }{
            \hspace{18pt}\sum_{i=1}^{n_{\mathrm{A}}}~
            \sum_{t=0}^{L[\textbf z^{(i)}_{ \mathrm{A}}]}
            \theta(p_{\mathrm{B}}(z^{(i)}_{\mathrm{A}}(t)) \geq \lambda_{\mathrm{A}})},
        \\
        \gamma_{\mathrm{B}}(\lambda_{\mathrm{B}}) &=
        \frac{
            \sum_{i=1}^{n_{\mathrm{TPS}}}
            w_{\mathrm{B}}^{(i)}~
            \sum_{t=0}^{L[\textbf x^{(i)}]}
            \theta(p_{\mathrm{B}}(x^{(i)}(t)) \leq \lambda_{\mathrm{B}})
        }{
            \hspace{18pt}\sum_{i=1}^{n_{\mathrm{B}}}~
            \sum_{t=0}^{L[\textbf z^{(i)}_{ \mathrm{B}}]}
            \theta(p_{\mathrm{B}}(z^{(i)}_{ \mathrm{B}}(t)) \leq \lambda_{\mathrm{B}})}.
    \end{align}
\end{subequations}
$\gamma_{\mathrm{A}}$ and $\gamma_{\mathrm{B}}$ should be constant 
for all choices of $\lambda_{\mathrm{A}},~\lambda_{\mathrm{B}}$. In practice, they become inaccurate when $\lambda_{\mathrm{A}}$ and $\lambda_{\mathrm{B}}$ are too close to $\mathrm{A}$ and $\mathrm{B}$ (due to relatively large error in the committor estimate) or too close to the transition state (due to the inadequate equilibrium sampling). In particular, from Eq.~\eqref{crossing0} a trajectory leaving $\mathrm{A}$ reaches $\lambda_{\mathrm{A}}$ a factor $\lambda_{\mathrm{A}}^{-1}$ times more frequently than undergoing a transition.

It is convenient to determine the value of $\lambda_{\mathrm{A}}$ and $\lambda_{\mathrm{B}}$ by fixing the number of equilibrium configurations that go beyond those thresholds:
\begin{subequations}\label{equilibrium_population}
\begin{align}
    M_{\mathrm{A}} &= \sum_{i=1}^{n_{\mathrm{A}}}~
            \sum_{t=0}^{L[\textbf z^{(i)}_{ \mathrm{A}}]}
            \theta(p_{\mathrm{B}}(z^{(i)}_{\mathrm{A}}(t)) \geq \lambda_{\mathrm{A}}), \\
    M_{\mathrm{B}} &= \sum_{i=1}^{n_{\mathrm{B}}}~
            \sum_{t=0}^{L[\textbf z^{(i)}_{ \mathrm{B}}]}
            \theta(p_{\mathrm{B}}(z^{(i)}_{\mathrm{B}}(t)) \leq \lambda_{\mathrm{B}}).
\end{align}
\end{subequations}
In this way, $\lambda_{\mathrm{A}}=\lambda_{\mathrm{A}}(M_{\mathrm{A}})$ and $\lambda_{\mathrm{B}}=\lambda_{\mathrm{B}}(M_{\mathrm{B}})$ follow from inversion.
By setting $M_{\mathrm{A}}$ and $M_{\mathrm{B}}$, we ensure that enough equilibrium sampling contributes to the calculation of $\gamma_{\mathrm{A}}(\lambda_{\mathrm{A}}(M_{\mathrm{A}}))$ and $\gamma_{\mathrm{B}}(\lambda_{\mathrm{B}}(M_{\mathrm{B}}))$. We can then optimize $M_{\mathrm{A}}$ and $M_{\mathrm{B}}$ as the parameters returning the most robust $\gamma_{\mathrm{A}}$ and $\gamma_{\mathrm{B}}$ to small boundary changes (Figs.~\ref{fig:S3},~\ref{fig:S4}b). 
The weights of Eq.~\eqref{weightsA} are independent of the choice of $\lambda_{\mathrm{A}}$ and $\lambda_{\mathrm{B}}$ and thus are computed only once.

In general, it is always possible to match the distributions around the wells and in the transitions region by using WHAM  or analogous approaches \cite{kumar1992weighted,stelzl_dynamic_2017,ferguson_bayeswham_2017}.

Finally, we enforce global normalization by rescaling all weights such that they sum up to one. 

\subsection{\label{sec:theoryF}Free energy profiles along arbitrary variables}

We can project the PE and get the equilibrium distribution $\rho$ as a function of any set of collective variables $q~=~\{q_1,~\dots,~q_k\}$\cite{rogal2010reweighted,bolhuis2011relation}. Starting from Eq.~\eqref{eq:abstract_projection}, making the weights explicit, and using the density operator $\delta(q(x)-q')$:
\begin{equation}
\label{densities}
\begin{aligned}
    \rho(q') \propto \gamma_{\mathrm{A}} \sum_{i=1}^{n_{\mathrm{A}}}\sum_{t=0}^{L[\textbf z^{(i)}_{\mathrm{A}}]} \theta(p_{\mathrm{B}}(z^{(i)}_{ \mathrm{A}}(t))< \lambda_{\mathrm{A}})~\delta(q(z^{(i)}_{ \mathrm{A}}(t))-q')\\
    \hspace{12pt}+\sum_{i=1}^{n_{\mathrm{TPS}}} w_{\mathrm{A}}^{(i)}\sum_{t=0}^{L[\textbf x^{(i)}]} \theta(p_{\mathrm{B}}(x^{(i)}(t))\geq \lambda_{\mathrm{A}})~\delta(q(x^{(i)}(t))-q')\\
    \hspace{12pt}+\sum_{i=1}^{n_{\mathrm{TPS}}} w_{\mathrm{B}}^{(i)}\sum_{t=0}^{L[\textbf x^{(i)}]} \theta(p_{\mathrm{B}}(x^{(i)}(t))\leq \lambda_{\mathrm{B}})~\delta(q(x^{(i)}(t))-q')\\
    +\gamma_{\mathrm{B}}  \sum_{i=1}^{n_{\mathrm{B}}}\sum_{t=0}^{L[\textbf z^{(i)}_{\mathrm{B}}]} \theta(p_{\mathrm{B}}(z^{(i)}_{\mathrm{B}}(t)) > \lambda_{\mathrm{B}})~\delta(q(z^{(i)}_{\mathrm{B}}(t))-q').
\end{aligned}
\end{equation}
The corresponding free energy profile is $F(q)=-k_{\mathrm{B}}T~\log \rho(q)$ up to an additive constant. The free energy difference between $\mathrm{A}$ and $\mathrm{B}$ is
\begin{equation}
    \Delta F_{\mathrm{AB}} = F_{\mathrm{B}}-F_{\mathrm{A}} = k_{\mathrm{B}}T~\log \frac{\rho_{\mathrm{A}}}{\rho_{\mathrm{B}}},
\end{equation}
where $\rho_{\mathrm{A}}$ and $\rho_{\mathrm{B}}$ are the sum of $\rho(q)$ over all configurations in $\mathrm{A}$ and $\mathrm{B}$, respectively.

\subsection{\label{sec:theoryG}Rate constants}

For the reaction rate constants, we employ the Bayesian framework developed by Hummer\cite{hummer2004transition,best2005reaction}:
\begin{equation}\label{rates}
    \nu=\frac{2}{k_{\mathrm{AB}}^{-1} + k_{\mathrm{BA}}^{-1}} = \frac{\rho(\lambda)}{\rho_{\mathrm{TP}}(\lambda)}~\frac{2\lambda(1-\lambda)}{\langle t_{\mathrm{AB}}\rangle_{\mathrm{TP}}},
\end{equation}
where $k_{\mathrm{AB}}$ and $k_{\mathrm{BA}}$ are the $\mathrm A\rightarrow \mathrm B$ and $\mathrm B\rightarrow \mathrm A$ rates, respectively, and we project all densities  on the committor. The transition path density $\rho_{\mathrm{TP}}$ comes from the previous AIMMD run. $\langle t_{\mathrm{AB}}\rangle_{\mathrm{TP}}$ is the average duration of the TPs, and $2\lambda(1-\lambda)$ is the probability $P(\mathrm{TP}\mid \lambda)$ of an equilibrium trajectory crossing $\lambda$ to be reactive, as in Eq.~\eqref{pTP}.

Eq.~\eqref{rates} holds for any value of $\lambda$, although it produces more accurate results around the transition state ($\lambda=0.5$) due to a better estimate of the committor. The outcome is the quantity $\nu$ combining both $k_{\mathrm{AB}}$ and $k_{\mathrm{BA}}$; it is the inverse of the average mean first passage time for $\mathrm{A}\rightarrow\mathrm{B}$ and $\mathrm{B}\rightarrow\mathrm{A}$, also known as the average return time. The individual rate constants can be obtained from the following:
\begin{subequations}
\begin{align}
    k_{\mathrm{AB}} = &\frac{1 + e^{-\Delta F_{\mathrm{AB}}/k_{\mathrm{B}}T}}{2}~\nu,\\\label{ratesBA}
    k_{\mathrm{BA}} = &\frac{1 + e^{+\Delta F_{\mathrm{AB}}/k_{\mathrm{B}}T}}{2}~\nu.
\end{align}
\end{subequations}
We could have also estimated the rates multiplying the reactive fluxes through the interfaces defined by $\lambda_{\mathrm{A}}$ and $\lambda_{\mathrm{B}}$ with the expected crossing probabilities\cite{van2003novel}. However, we 
found that Eq.~\eqref{rates} provides a numerically more robust estimate. 

\subsection{\label{sec:theoryE}The complete algorithm}

We summarize the entire algorithm:

\begin{itemize}
    \item[1.] Perform AIMMD simulations and learn the committor $p_{\mathrm{B}}(x)$ from the $\{x_{\mathrm{sp}}^{(i)},~r^{(i)}\}$ training set. Collect the $\textbf x^{(1)},~\textbf x^{(2)},~\dots,~\textbf x^{(n)}$ trial paths (note that this includes the rejected paths).
    \item[2.] At the same time, run MD from multiple equilibrium configurations in states $\mathrm{A}$ and $\mathrm{B}$. Collect the sampled $\textbf z^{(i)}_{\mathrm{A}}$ and $\textbf z^{(i)}_{\mathrm{B}}$ trajectories.
    \item[3.] Evaluate $p_{\mathrm{B}}$ on all the simulated configurations; for each $\textbf x^{(i)}$, save $\lambda^{(i)}=p_{\mathrm{B}}(x^{(i)}_{\mathrm{sp}}),~\lambda_{\min}^{(i)}$, and $~\lambda_{\max}^{(i)}$.
    \item[4.] Weight the $\textbf x^{(i)}$ for approximating $\mathcal P_{\mathrm{A}, \lambda_{\mathrm{A}}}^+$ and $\mathcal P_{\mathrm{B}, \lambda_{\mathrm{B}}}^-$ according to  Eqs.~\eqref{weights}, with preliminary $c_{\mathrm{A}}=c_{\mathrm{B}}=1$. Obtain the unnormalized $w_{\mathrm{A}}^{(i)}$ and $w_{\mathrm{B}}^{(i)}$ (Fig.~\ref{fig:2}d).
    \item[5.] Evaluate $\gamma_{\mathrm{A}}$ and $\gamma_{\mathrm{B}}$ from Eq.~\eqref{gamma} and determine the optimal $\lambda_{\mathrm{A}}$ and $\lambda_{\mathrm{B}}$ parameters. Assign the weights $\textbf v_{\mathrm{A}}^{(i)}$ and the $\textbf v_{\mathrm{B}}^{(i)}$ trajectories according to $\gamma_{\mathrm{A}}(\lambda_{\mathrm{A}})$ and $\gamma_{\mathrm{B}}(\lambda_{\mathrm{B}})$.
    \item[6.] Project $\mathcal P_{\mathrm{A}}$ and $\mathcal P_{\mathrm{B}}$ on the transition state: $\rho_{\mathrm{A}}(\lambda=0.5)$ and $\rho_{\mathrm{B}}(\lambda=0.5)$. Impose the condition of Eq.~\eqref{normalization} by rescaling $c_{\mathrm{A}}=1/\rho_{\mathrm{A}}(0.5)$ and $c_{\mathrm{B}}=1/\rho_{\mathrm{B}}(0.5)$, and thus $w_{\mathrm A}^{(i)}$, $w_{\mathrm B}^{(i)}$, $\gamma_{\mathrm A}$, and $\gamma_{\mathrm B}$. 
    \item[7.] Merge all the simulated trajectories as in Eq.~\eqref{ensemble}. Normalize the weights over all configurations.
    \item[8.] Obtain a free energy profile as a function of the estimated committor, or as a function of arbitrary variables, and estimate the rate constants. 
\end{itemize}

\section{\label{sec:methods}Methods}

\subsection{\label{sec:methodsA}AIMMD and PE calculations}
% Alternative title "Software"

We used and extended the AIMMD Python package developed by Jung\cite{jung2023machine} to run the path sampling simulations.
For each AIMMD run, we initialized a deep neural network in PyTorch\cite{paszke2019pytorch} with 4 hidden linear layers of size 8192, 2048, 512, and 128 with ELU activation functions, 4 residual units\cite{kaiming2016identity} with 4 layers and 128 neurons per layer, and a final linear layer (Fig.~\ref{fig:S1}). The architecture is expressive enough to capture the shape of the committor in many-dimensional systems. The encoder structure encourages the pruning of unimportant features.

The network's output is $q(x)$, which is a one-to-one function of the committor \cite{daqi2005classification}:
\begin{equation}
    p_{\mathrm{B}}(x) = \sigma(q(x)) = \frac{1}{1+e^{-q(x)}}.
\end{equation}

After every TPS step, we reset and trained the network on all the available shooting points accumulated up to that point. We performed 100 training epochs by minimizing the binomial loss of Eq.~\eqref{loss} with the ADAM optimizer\cite{diederik2014adam}. We found that the learning rates $l_r=10^{-5}$ (2D systems) and $l_r=3.5\times 10^{-6}$ (chignolin), dependent on the network architecture, were good hyperparameter choices to prevent the model from overfitting\cite{ying2019overview}. We observed that the optimal $l_r$ value decreased with increased input feature dimensionality, with no system-specific dependency. To regularize the training set, we set the shooting points' importance $v^{(i)}$ such that the A-to-A, the B-to-B, and the A-to-B and B-to-A results would have each equal cumulative weight. We saved the neural network parameters at regular intervals.

To achieve the target uniform selection probability, we determined $p_{\mathrm{sel}}$ based on the committor values of the origin trajectory (the last accepted trajectory in the Markov chain built by TPS). We organized the candidate shooting points into 10 equally $p_{\mathrm{B}}$-spaced bins and scaled the probability by the bins' population. The probability of an empty bin was  distributed to the adjacent ones. For each trajectory, the selection probability is a function of the committor only. To ensure detailed balance in the Markov chain, we kept the rule consistent within a step when calculating the acceptance probability of Eq.~\eqref{acceptance}.

We wrote the PathEnsemble Python class to automate the PE estimation summarized in Section~\ref{sec:theoryE}. A PathEnsemble instance collects the features vectors, committor values, and complementary information of a set of trajectories. It can extract the TPE from the trials of a TPS run, weight the trajectories for estimating $\mathcal P_{\mathrm{A}}$ and $\mathcal P_{\mathrm{B}}$, and combine ensembles together. It can also project the free energy onto an arbitrary set of collective variables. 

\subsection{\label{sec:methodsC}2D systems}

\begin{figure*}
    \centering
    \includegraphics[width=\textwidth]{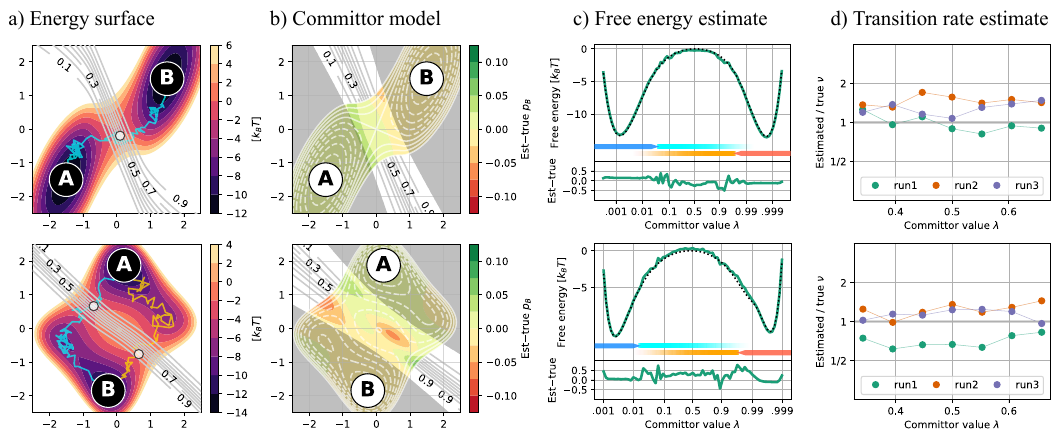}
    \caption{Validation on the two-dimensional benchmark systems (top: double well, bottom: Wolfe-Quapp), results after 500 AIMMD steps. \textbf{a)} Potential energy surface (filled contour), example of TPs in different channels (blue and orange lines, shooting points in white), and true committor (contour lines). \textbf{b)} AIMMD run1, committor model (contour lines), error of the model (filled contour), and region between the $\lambda_{\mathrm{A}}$ and $\lambda_{\mathrm{B}}$ thresholds (light area). \textbf{c)} Run1, free energy as a function of the estimated committor (solid line, top axis), free energy from numerical computation (dotted line), and error of the estimate (bottom axis). The arrows indicate the contributions of the simulations around A (blue), the TPS trajectories from A (cyan), the simulations around B (red), and the TPS trajectories from B (orange). \textbf{d)} Bayesian rate estimate of $\nu$ at different committor values, each color denoting a different run. The gray area is the 95\% confidence interval of $\nu$ from   equilibrium simulations.}
    \label{fig:3}
\end{figure*}

The two-dimensional (2D) systems are defined by their  energy surface on the $(x,~y)$ plane. The double well energy surface has equation:
\begin{equation}
U(x, y) =
\begin{cases}
-2~\Delta G~(x/\delta)^2 + k_0~(x-y)^2/2\\
\hspace{50pt}\text{if  } x/\delta < 0.5,\vspace{3pt}\\
\Delta G~[2 (|x/\delta| - 1)^2 -1] +k_0~(x-y)^2/2\\
\hspace{50pt}\text{if  } x/\delta \geq 0.5\text{;}
\end{cases}
\end{equation}
with the barrier height $\Delta G = 12\ k_{\mathrm{B}}T$, $k_0 = 10.4\ k_{\mathrm{B}}T$, and $\delta = 1.5$ (Fig.~\ref{fig:3}a, top). The Wolfe-Quapp (Wolfe-Quapp) energy surface\cite{quapp2015growing} has equation:
\begin{equation}
U(x, y) = \frac{\Delta G}{5}~(x^4 + y^2 - 2x^2 - 3y^2 + xy + 0.3x + 0.1y)\text{;}
\end{equation}
we set $\Delta G = 10\ k_{\mathrm{B}}T$, and rotated the $x$ and $y$ coordinates by 45 degrees (Fig.~\ref{fig:3}a, bottom). 

In each system, we evolved a particle with overdamped Langevin dynamics\cite{peters2017reaction}
(diffusion coefficient $D=10^{-5}$ with unitary distance, energy, integration time step, and mass), and saved the trajectories every 500 (double well) and 1,000 (Wolfe-Quapp) integration steps.
In this way, TPs will contain approximately $100$ frames. As metastable states, we picked circles of radius $r=0.5$ around the local minima. We computed the reference committor by numerically solving the stationary Fokker-Plank equation\cite{covino_molecular_2019} and derived the reference $k_{\mathrm{AB}}$ and $k_{\mathrm{BA}}$ rate constants by fitting the exponential decay\cite{peters2017reaction} of 40,000 replicas initiated in A and B.

For each system, we performed 3 AIMMD runs of 5,000 steps each, directly feeding the $x,y$ coordinates to three different neural networks. To assess the speed-up given by AIMMD, we performed a standard TPS (run0) as a benchmark. As initial trajectory (then excluded from the TPE), we drew a straight line connecting the minima; the first trial TP is always accepted.

From each AIMMD run, we computed and extended the PE with 20 equilibrium trajectories initiated from the energy minima (10 each). The trajectories are at most $500,000$ (double well) and $25,000$ (Wolfe-Quapp) frames long and were trimmed in case they reached the other state. We determined the $\lambda_{\mathrm{A}},~\lambda_{\mathrm{B}}$ thresholds according to Eq.~\eqref{equilibrium_population} such that 100 configurations from A and from B went beyond those committor values (Fig.~\ref{fig:S3}). For the ideal scenario of optimal sampling around the basins, we numerically computed the $\rho_{\mathrm{A}}(x, y)=\rho(x,y)~(1-p_{\mathrm{B}}(x,y))$ and $\rho_{\mathrm{B}}(x, y)=\rho(x,y)~p_{\mathrm{B}}(x,y)$ distributions, and scaled their weights such that $\lambda_{\mathrm{A}},~\lambda_{\mathrm{B}}$ were consistent with the previous case.

\subsection{\label{sec:methodsD}Chignolin}

We obtained the folded structure of CLN025 (amino acid sequence YYDPETGTWY) from the 2RVD entry of Protein Data Bank\cite{kato2015nmr,yasuda2014physical} (Fig.~\ref{fig:5}b). We solvated the peptide with TIP3 water in a $4$ nm cubic periodic box and generated a topology file with Charmm-GUI\cite{sunhwan2008charmmgui}; the final system has 6,468 atoms, 166 belonging to the peptide. We reproduced the settings of Lindorff-Larsen et al.\cite{lindorff2011fastfolding} and chose the CHARMM22$^\star$ force-field\cite{piana2011robust}. We ran the simulations with GROMACS 2022.4\cite{bauer2022ym} and the velocity Verlet integrator; we fixed the volume after $1$ ns of equilibration and kept the temperature $T=340$ K with the velocity rescale thermostat\cite{bussi2007canonical}. We set a $0.95$ nm threshold for the short-range interactions  and left the remaining GROMACS  parameters unchanged.
We integrated the positions and momenta every $dt=2$ fs and saved the former every $\Delta t=100$ ps in XTC trajectory files.

We calculated the reference free energy profiles and rate constants from 4 equilibrium MD simulations, totaling $120~$µs. We visualized the trajectories with VMD\cite{humphrey1996vmd} and analyzed them with MDTraj\cite{mcgibbon2015mdtraj}.

We defined the folded ($\mathrm{A}$) and unfolded ($\mathrm{B}$) states based on the fraction of native contacts\cite{best2013native}:
\begin{subequations}\label{boundaries}
\begin{align}
    \mathrm{A} &= \{ x \mid Q(x) \geq 0.99\},\\
    \mathrm{B} &= \{ x \mid Q(x) \leq 0.01\},
\end{align}
\end{subequations}
where the reference configuration ($t=55.1$ ns of the first equilibrium MD simulation) is the centroid of the $\mathrm{C}_\alpha$-RMSD folded state cluster\cite{lindorff2011fastfolding}. 
We also considered the following additional collective variables (see also Figure~\ref{fig:6}):
\begin{itemize}
    \item[1.] the distance between Asp3O and Gly7N ($d_1$), forming a hydrogen bond in the native state\cite{satoh2006folding};
    \item[2.] the distance between Asp3N and Gly7O ($d_2$);
    \item[3.] the distance between Asp3N and Thr8O ($d_3$);
    \item[4.] the fraction of native contacts between Tyr2 and Trp9 ($Q_{29}$)\cite{harada2011exploring};
    \item[5.] the fraction of native contacts between Pro4 and Gly7  ($Q_{47}$);
    \item[6.] the C$\alpha$-RMSD with respect to the reference structure\cite{lindorff2011fastfolding};
    \item[7.] the radius of gyration of the protein's heavy atoms ($r_g$).
\end{itemize}

We performed 3 AIMMD runs of 250 steps each. As in the 2D systems, we ran further standard TPS (run0, run0b, and run0c) with random selection probability as a benchmark to compare performances.
As the input for the neural network, we featurized the system calculating 2064 distances between heavy atoms at least 4 residues apart, and min-max normalized the distances according to the values sampled in the short equilibrium simulations in $\mathrm{A}$ and $\mathrm{B}$. 

To obtain the initial trajectory (then excluded from the TPE), we raised the temperature to $T=600~\text{K}$ and quickly unfolded the folded state in $0.9~\text{ns}$; the first reactive trial path is always accepted. In extending the PE, we integrated each AIMMD run with 20 short simulations initiated from two original 20 ns equilibrium trajectories around A and B (Fig.~\ref{fig:2}b). The simulations were terminated as soon as they hit 50 ns (500 frames, for A) or 5 ns (for B). We picked the $\lambda_{\mathrm{A}}$ and $\lambda_{\mathrm{B}}$ thresholds such that 10 configurations around A and 50 around B crossed those committor values (Fig.~\ref{fig:S4}b).

\section{\label{sec:results}Results and discussion}

\subsection{\label{sec:resultsA}Double well potential}

We illustrated our method on the double well benchmark system. The potential has a $12~k_{\mathrm{B}}T$ energy barrier, resulting in a mean first passage time about $10^5$ times larger than the average TP time. The committor varies significantly in a small portion of the transition region, with most equilibrium configurations highly committed to either $\mathrm{A}$ or $\mathrm{B}$ (Fig.~\ref{fig:3}a, top). Many configurations of the TPE are also far from the barrier. Hence biasing the shooting point selection probability towards the transition state is essential for good sampling performance\cite{falkner2022conditioning}.

AIMMD generated 4,698 TPs and 2,456 accepted ones in 15,000 steps across 3 independent runs. The resulting TPEs match the reference (Fig.~\ref{fig:S2}a).
% it soon improved from the standard TPS
To mimic a data-poor regime, we calculated committor, free energy, and rates using only the first 500 steps, corresponding to 163 TPs.
The networks quickly learned the committor (Fig.~\ref{fig:3}b, top row), with an absolute error of $p_{\mathrm{B}}(x)$ below $0.05$ in the reactive channel.

We computed the PE from the AIMMD run1 data and projected the free energy on the committor estimated by the network (Fig.~\ref{fig:3}c, top). 
The absolute error of $F(q)$ remains below $0.2~k_{\mathrm{B}}T$ once aligned to the target. 

We estimated the kinetics of the system by calculating $\nu$ with the Bayesian approach of Eq.~\eqref{rates} at different committor values $\lambda$. Again, we stopped at 500 TPS steps and plotted the results of all the 3 runs to show the statistics (Fig.~\ref{fig:3}d, top). The estimates from the same run are stable between the $\lambda_{\mathrm{A}}$ and $\lambda_{\mathrm{B}}$ thresholds (light area in Fig.~\ref{fig:3}b). They range between $0.8$ and $1.5$ times the reference rate. Each run took about $0.0035$ of cumulative simulations in $\nu^{-1}$ units. By also adding the sampling around the states for the PE extension, the total simulated time reaches $0.742~\nu^{-1}=0.742~k_{\mathrm{AB}}^{-1}$. We expect no transitions at all from an equilibrium simulation of the same length. In contrast, our method successfully provided accurate free energy, rates, and also learned the reaction coordinate for the transition.

\subsection{\label{sec:resultsB}Wolfe-Quapp potential}

\begin{figure}
\centering
\hspace{-4.5pt}
\includegraphics[width=\columnwidth]{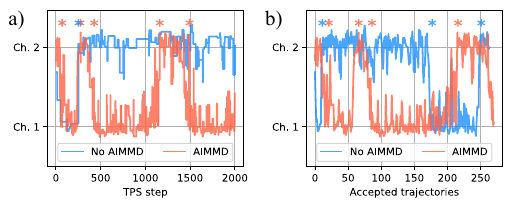}
    \caption{Reactive channel switch during AIMMD simulations on the Wolfe-Quapp energy surface. The plots show the channel containing the last accepted TP with the number of TPS steps \textbf{(a)} or accepted moves \textbf{(b)}. The red lines represent AIMMD run1, while the blue lines are from the standard (uniform selection) TPS run0. The channel is determined by computing the Hausdorff distance to a reference path in Channel 1. The stars mark the channel switches.}
    \label{fig:4}
\end{figure}

The Wolfe-Quapp potential (Fig.~\ref{fig:3}, bottom row) contains two alternative reaction channels posing an additional challenge for the sampling. The two channels have different energy profiles and travel times across the isocommittor surfaces. Switching channels requires crossing a separation barrier of $2~k_{\mathrm{B}}T$ at the transition state.

AIMMD substantially increased the switches frequency compared to standard TPS both  considering the total steps and the accepted trajectories alone (Fig.~\ref{fig:4}). As a result, it took 400 steps on average to switch between channels. Occasionally selecting shooting points close to the states helped, as that promoted exploration of different configurations with a reasonable toll on exploitation: at convergence, the expected TPs' production rate ($0.33$) is 66\% the theoretical maximum (0.5, when selecting shooting points only at the transition state). We stress that we only used the instantaneous $p_{\mathrm{B}}$ in determining the selection probability. One could tune the exploration-exploitation trade-off by shaping $p_{\mathrm{sel}}(p_{\mathrm{B}})$, although we found the uniform solution optimal for improving both the committor mode and the estimated PE accuracy. A well-tailored selection bias would help further decrease the path decorrelation.

We put ourselves in a data-poor regime and took the first 500 steps (and 172 TPs) of AIMMD run1. The resulting committor is less accurate in the low-energy channel (Fig.~\ref{fig:3}b, bottom), albeit on par with the double well system overall. The speed boost of AIMMD allowed exploring both channels; however, we put ourselves in a data-poor condition where their relative importance is hard to infer. This especially affects the evaluated TPE (Figs.~\ref{fig:S2}b). Notwithstanding, we obtained an excellent estimate of the free energy profile as a function of the estimated committor, and the estimated rates remain within a twofold error between $\lambda_{\mathrm{A}}$ and $\lambda_{\mathrm{B}}$ (Fig.~\ref{fig:3}d, bottom). In each run, we simulated approximately $0.071$ of cumulative time in $\nu^{-1}$ units, $0.722~\nu^{-1}$ by including the additional sampling in the metastable states.

 \begin{figure}
    \centering
    \hspace{-4.5pt}
    \includegraphics[width=\columnwidth]{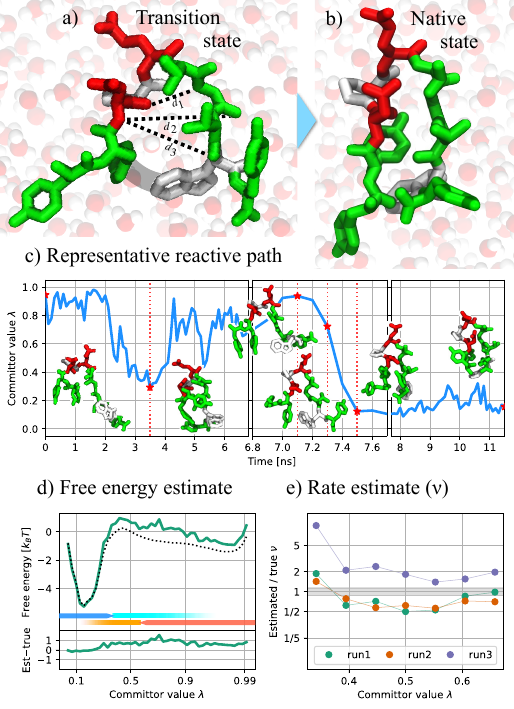}
    \caption{Characterizing the folding of chignolin. \textbf{a)} Chignolin's transition state configuration, no-hydrogen licorice representation colored in white (non-polar residues), red (acidic), and green (polar). We highlighted the formation of the $d_1$, $d_2$, and $d_3$ hydrogen bonds and the association of Tyr2 and Trp9.
    \textbf{b)} Chignolin's native structure represented as in \textbf{a}.
    \textbf{c)} Representative folding trajectory (step 100 of run1), committor time series with renders of highlighted configurations. The model was trained on the first 50  steps of run1.
    \textbf{d)} AIMMD run1, free energy as a function of the committor after 50 steps (solid line, top axis), free energy from long equilibrium simulations (dotted line), and difference between the two (bottom axis). The arrows indicate the contributions of the simulations around A (blue), the TPS trajectories from A (cyan), the simulations around B (red), and the TPS trajectories from B (orange).
    \textbf{e)} Bayesian rate estimate of $\nu$ at different committor values, each color denoting a different run. The gray area is the 95\% confidence interval of $\nu$ from  long equilibrium simulations.}
    \label{fig:5}
\end{figure}

\subsection{\label{sec:resultsC}Chignolin}

 \begin{figure}
    \centering
    \includegraphics[width=\columnwidth]{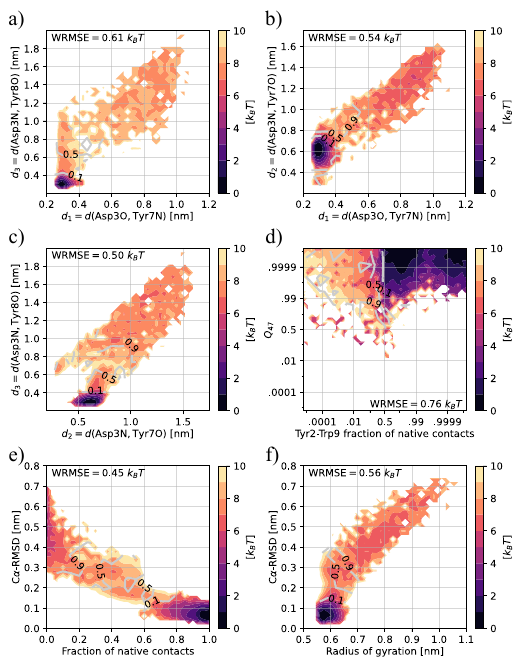}
    \caption{Chignolin's free energy profiles (colored contour) and effective committor (contour lines) extracted from the PE estimated from the first 250 steps of AIMMD run1, projected on 6 pairs of collective variables. We annotated the root mean square deviation $\sqrt{\left<(F - \tilde F) ^ 2\right>_{\rho}}$ from the reference free energy $\tilde F$ obtained from long equilibrium MD simulations: .}
    \label{fig:6}
\end{figure}

Chignolin is the smallest peptide folding into a $\beta$-hairpin structure\cite{yasuda2014physical}. The CLN025 variant exhibits a two-state behavior with the folded state showing remarkable stability\cite{davis2012raising,sumi2019theoretical} (Fig.~\ref{fig:5}b). The small size and short transition time, together with the formation of non-local structures, make it a good molecular system for testing our method and an entry point to studying more complex molecular events. From our equilibrium MD simulations, we estimated a folding rate $k_{\mathrm{BA}} = 2.5(5)~\text{µs}^{-1}$, an unfolding rate $k_{\mathrm{AB}} = 0.28(5)~\text{µs}^{-1}$, a combined $\nu=0.50(7)~\text{µs}^{-1}$, and a transition path time $\langle t_{\mathrm{AB}} \rangle_{\mathrm{TP}} = 11(2)$ ns, in agreement with Lindorff-Larsen et al\cite{lindorff2011fastfolding}. 

AIMMD provided accurate mechanism, free energy and rates of folding with only a handful of trajectories. We focus on AIMMD run1 (the other two runs yielded consistent results, see also Fig.~\ref{fig:S5}). After only 50 steps (containing 20 TPs), the committor clearly distinguishes between the folded and unfolded state (Fig.~\ref{fig:5}c) and is accurate when validated on an independent data set (Fig.~\ref{fig:S4}a). This limited number of trajectories produces a folding free energy profile within $1~k_{\mathrm{B}}T$ of the expected value at the barrier (Fig.~\ref{fig:5}d, see also Fig.~\ref{fig:S4}c for the individual contributions to the PE). Also, we could estimate $\nu=0.25~\text{µs}^{-1}$, which is less than a factor 2 away from the reference value. The folding rate $k_{\mathrm{BA}}=0.85~\text{µs}^{-1}$ is compatible with the estimate from very long equilibrium simulations (Fig.~\ref{fig:S4}d); the two other runs were less accurate, but still within an order of magnitude from the reference. The run took $1.10$ µs of cumulative simulated time, corresponding to $0.548~\nu^{-1}$. When extending the estimate of the PE to the metastable states, we chose the largest $M_{\mathrm{A}}$ and $M_{\mathrm{B}}$ that ensured the stability of the reweighting factors (Fig.~\ref{fig:S4}b). We emphasize that the simulations used for the PE extension are short and confined to the states. The TPS trajectories are thus essential to combine the two equilibrium path ensembles associated with each state with the proper weights. The additional data added up to $0.90~\text{µs}$, or $0.448~\nu^{-1}$. 
 
Our method enables the characterization of the folding mechanism beyond free energy and rates. One could directly inspect the TPs or obtain explicit models of the committor with AIMMD (Fig.~\ref{fig:5}c). Another way is to project the estimated PE on selected collective variables through Eq.~\eqref{densities}. This has the advantage of allowing the expert to choose among standard domain-specific features and comes with no extra computational cost. It also produces a multifaceted representation of the process, putting the accent on different aspects and therefore rendering a more complete picture of this complex re-organization. We obtained the free energy on 6 pairs of collective variables discussed in the literature\cite{satoh2006folding,harada2011exploring} (Fig.~\ref{fig:6}), all in good agreement with the equilibrium MD data.

Another advantage of estimating the PE is that it provides the effective (generalized) committor\cite{bolhuis2011relation} in any reduced space through Eq.~\eqref{effective_committor}. $p_{\mathrm{B}}$ encodes the progress of the reaction and complements the free energy information. For example, Fig.~\ref{fig:6}d reveals alternative pathways to the ``turn zipper'' folding mechanism. Here, the Tyr2-Trp9 contacts, independent of the complete formation of Pro4-Gly7, are the real limiting factor in the reaction.
Similar behavior was already reported in the literature and associated with hydrophobic collapse\cite{davis2012raising,dinner1999understanding}.
The Asp3,N-Tyr8,O H-bond formation\cite{mckiernan2017modeling} is another crucial event at the barrier (Fig.~\ref{fig:6}a): this, along with the fraction of native contacts and the C$\alpha$-RMSD, stand out as the most important features in separating the folded and unfolded state.

\begin{figure*}
    \centering
    \includegraphics[width=\textwidth]{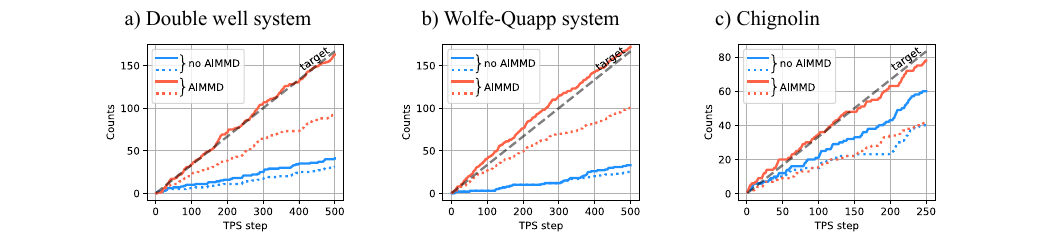}
    \caption{Efficiency of generating TPs by AIMMD compared to standard TPS, for the double well (\textbf{a}), Wolfe-Quapp (\textbf{b}) and chignolin (\textbf{c}). The plots show the cumulative number of generated TPs (solid lines) and accepted ones (dotted lines). The red lines denote AIMMD run1, while the blue lines are from TPS run0. The dashed line (target) is the optimal upper-bound number of TPs given the chosen AIMMD selection probability.}
    \label{fig:7}
\end{figure*}

\begin{figure*}
    \centering
    \includegraphics[width=\textwidth]{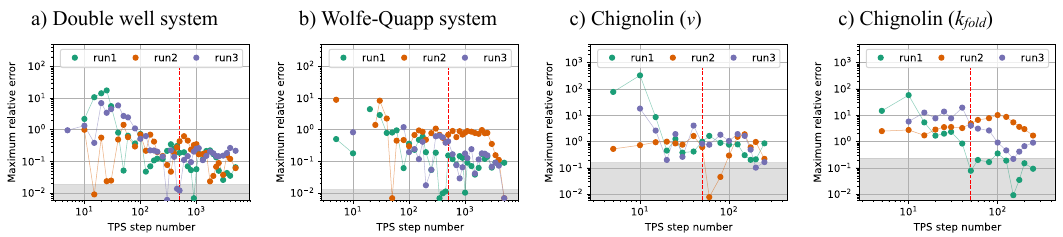}
    \caption{Accuracy of estimated kinetics as a function of an increasing number of simulations for the double well (\textbf{a}), Wolfe-Quapp (\textbf{b}) and chignolin (\textbf{c}). The plot shows the maximum relative error of the $\nu$ estimate at $\lambda=0.5$ (transition state) as a function of the number of TPS steps. Each color corresponds to a different AIMMD run. For chignolin, we also plot the error of the folding rate $k_{\mathrm{BA}}$ (\textbf{d}). The vertical lines mark the steps considered in Figs.~\ref{fig:3}~and~\ref{fig:5}. The gray area is the 95\% confidence interval of $\nu$ from equilibrium simulations.}
    \label{fig:8}
\end{figure*}

\subsection{\label{sec:resultsD}Performance evaluation}

We assessed the performance of our proposed algorithm under data-poor and rich regimes, both in terms of required computational resources and the quality of our estimates. In the data-poor scenario, we restricted the total simulations below the (average) mean first passage time. In the data-rich situation, we imposed no limitation on the computational resources to evaluate the highest expected accuracy achievable by our method. We stress that all the estimates presented until now fall in the data-poor regime. The results by including the complete simulations are collected in Figs.~\ref{fig:S6},~\ref{fig:S7},~\ref{fig:S8}.

We verified AIMMD's capability of accelerating TPS from its early stages (Fig.~\ref{fig:7}). The gain in production depends on how quickly the network converges to a reasonable committor model and how peaked the TPE density is at the transition state. All the systems promptly increased their production rate of TPs. Chignolin's higher chance of generating TPs from random configurations resulted in no significant difference in the number of accepted trajectories; in general, the system's complexity led to a higher variability within and among the runs. However, the most significant advantage of AIMMD is in obtaining the transition mechanism, free energies, and rates. To achieve this objective, learning the committor is crucial. Notably, training the network \textit{a posteriori} on standard TPS results led to worse committor models than using AIMMD (Fig.~\ref{fig:S9}).

To evaluate the gain of the full procedure, we focused on the rate estimates as the most illustrative example since they require evaluating the PE and TPE across the entire configuration space; plus, they are notably difficult to obtain with state-of-the-art techniques. To assess the computational resources, we also considered the total  simulated time in units of $\nu^{-1}$ but excluded the equilibrium simulations around the states as they can be executed in an embarrassingly parallel way while doing path sampling. In the 2D systems, we replaced the simulations in the states with the reference $\rho_{\mathrm{A}}$ and $\rho_{\mathrm{B}}$ to isolate the error arising from the underlying approximations of our method and see how accuracy scales with  sampling. Training the networks on GPU and estimating the PE took a negligible fraction of the resources dedicated to MD.

At every stage of the AIMMD runs, the accuracy of the rate estimates consistently outperformed the predictions from equilibrium simulations of matching duration (Fig.~\ref{fig:8}). This is especially true in the data-poor scenario when the simulated time is less than $\nu^{-1}$ and no spontaneous transitions are expected to happen. In all the systems, the relative error quickly dropped to a factor 2 after a few TPs and consistently reduced up to 10\% with the increasing number of TPS steps. %Chignolin's folding rates are less precise but still on point. 
At convergence, a small systematic error emerges from the discrete time interval between trajectory points. In particular, the true $\lambda_{\max}$ is always bigger than the recorded one for an excursion from $\mathrm{A}$, slightly altering the free energy profiles (Fig.~\ref{fig:S6}b).
When considering the total simulated time (Fig.~\ref{fig:S10}), the performance gain depends on the factor $\eta=\nu\cdot \langle t_{\mathrm{AB}}\rangle_{\mathrm{TP}}$ (the ``rareness'' of the event). Despite longer TPs, AIMMD enabled a significant computational gain applied to the study of chignolin's folding, also yielding reasonable folding rates (Fig.~\ref{fig:8}d). Moreover, the two competing pathways did not compromise the results in the Wolfe-Quapp system, even in the case of no channel switches throughout the simulations.

Remarkably, substituting the instantaneous committor with its numerical computation did not significantly improve the rate estimate for the 2D systems aside from the early TPS steps (Fig.~\ref{fig:S11}). We believe this is due to the robustness of the Bayesian approach while accounting for the ensembles' fluctuations. In particular, deviations in the TPE density $\rho_{\mathrm{TP}}(\lambda)$ are likely to reflect on $\rho(\lambda)$ and counterbalance in Eq.~\ref{rates}. Although it is possible to apply the method on TPS data trained \textit{a posteriori}, the estimate is generally worse (Fig.~\ref{fig:S11}). Finally, the adaptive $\lambda_{\mathrm{A}}$ and $\lambda_{\mathrm{B}}$ thresholds allowed for accurate results when the network underfits the committor close to the states.

\section{\label{sec:conclusions}Discussion and Conclusions}

Understanding how molecules dynamically organize is key to revealing how they function and enabling technological and biomedical breakthroughs. This understanding comes in two ways: an accurate quantitative description and a qualitative explanation that allows us to obtain an intuitive insight and paves the way to formulating hypotheses and models. MD simulations can provide both—the first as free energies and rates and the second as mechanisms. However, standard simulation schemes usually cannot sample the timescales required for either goal. 

Here, we presented a path sampling algorithm that gives access at the same time to mechanisms, free energy, and rates for rare events in molecular systems. Our algorithm is general, straightforward, and produces good free energy and rate estimates at a moderate computational cost. In essence, we run AIMMD simulations \cite{jung2023machine} to sample trajectories that explore the transition region between two metastable states, which we then reweight and integrate with short, unbiased simulations in the states. Our algorithm provides a free energy profile that can be projected on any collective variable. The only requirement of our algorithm is a definition of the two states and an initial trajectory connecting them. AIMMD will adaptively learn how to simulate optimally TPs and learn the committor. The simulations in the basins are simple, unbiased simulations that can be run in parallel to the path sampling. 

Our algorithm builds on established path sampling approaches and overcomes some of their limitations. TIS is a powerful method to obtain rates by seeding paths at interfaces between two states. While able to yield very accurate rate estimates, TIS is computationally expensive. Here, we combined a path recycling scheme using straightforward TPS two-way shooting simulations \cite{brotzakis2019approximating} with the RPE theory \cite{rogal2010reweighted} to approximate the equilibrium path distribution between the states. 
From a more abstract viewpoint, next to a regular Markov chain sampling from the constrained TP distribution, our method  creates a set of trajectories containing all TPS trial paths, which are all acceptable in the  equilibrium path ensemble. Giving each path in this set the correct weight yields an approximation of the equilibrium PE.
Our method thus performs regular TPS,  but takes advantage of the trial paths in an unorthodox way, setting it apart from other path sampling approaches.

The high efficiency of our algorithm relies on two factors: AIMMD samples TPs with near-to-optimal efficiency; the committor model learned by AIMMD is the ideal reaction coordinate that simplifies the reweighting  and makes it numerically more robust. Using the committor in combination with straightforward two-way-shooting TPS simulations radically simplifies the algorithm in practice and enables us to recycle existing TPS simulation campaigns to extract free energy and rates a posteriori.  

Rates are among the most challenging quantities to estimate in MD simulations. While many techniques exist to evaluate free energy profiles, rate calculations are much less established. On the other hand, free energy profiles are not observables---only free energy differences between metastable states are---while rates can often be measured in experiments, providing a natural way of comparing experiments and simulations. We anticipate that our algorithm and analogous approaches\cite{palacio-rodriguez_free_2022} will make the calculation of rates from MD simulation more accessible. By comparing calculated and measured rates, we can assess the systematic uncertainties arising from using semi-empirical force fields, which generally were not parameterized on kinetic measurements. 

Despite the many advancements, important challenges remain. Our algorithm focuses on characterizing rare molecular events between two states. While AIMMD and the underlying committor theory generalize to transition between multiple states\cite{rogal2008multiple,jung2023machine}, in practice, it might be more efficient to reduce this problem to a collection of pairwise transitions. The definition of two states is not always straightforward. However, it requires only order parameters that do not have to resolve the transition. Also, state definitions can be iteratively refined by using the committor. After a first simulation campaign, configurations with committor values close to 0 and 1 can be used as new, more accurate state boundaries. The correlation along the chain of sampled TPs is still a great challenge \cite{ghamari_sampling_2022}. We showed how AIMMD alleviates this problem by speeding up the switching between alternative reactive channels, but many steps are still required. Integrating generative AI approaches, as recently proposed by Dellago and coworkers, might provide the solution \cite{falkner2022conditioning,falkner2023enhanced}.

Our algorithm relies on many short, unbiased simulations. The clear advantage is that the dynamics are not distorted, and the reweighting is necessary only to obtain the correct stationary distribution in the transition region. This also means that the longest timescale that one must be able to simulate is the duration of TPs. These are usually exponentially shorter than the typical lifetime in the states and can be on the scales of nanoseconds, even for large and complex systems\cite{okazaki_mechanism_2019}. However, it will be challenging for some systems to sample a few TPs in a reasonable time. In addition, the energy wells of the states could be so deep that sampling excursions that overlap with the transition region could be impractical. Using a static biasing potential can help in both cases\cite{henin2022enhanced}. 

Our algorithm is simple to use and data-efficient. It builds on highly efficient simulation packages like GROMACS\cite{abraham2015gromacs} and OpenMM\cite{eastman2017openmm}. In this way, it seamlessly capitalizes on new software and conventional force-field developments. But it will also take advantage of the latest exciting developments in generative AI for conformational sampling\cite{noe2019boltzmann}. Path sampling simulations are increasingly more attractive for investigating rare molecular events.

\begin{acknowledgments}

We thank Drs. Jutta Rogal and Attila Szabo for stimulating discussions and helpful comments.
G.L. and R.C. acknowledge the support of the Frankfurt Institute of Advanced Studies, the LOEWE Center for Multiscale Modelling in Life Sciences of the state of Hesse, the CRC 1507: Membrane-associated Protein Assemblies, Machineries, and Supercomplexes, and computational resources and support by the SURFsara National Supercomputing and e-Science Support Center in The Netherlands, the Center for Scientific Computing of the Goethe University, and the Jülich Supercomputing Centre. G.L. was supported by a grant from the HPC-Europa3 program and acknowledges support of the iQbio graduate school of the Goethe University. R.C. acknowledges the support of the International Max Planck Research School on Cellular Biophysics. H.J. acknowledges support by the Max Planck Society. 
\end{acknowledgments}

\section*{Data Availability Statement}

For the purpose of Open Access, the author has applied a CC-BY license to any Author Accepted Manuscript version arising from this submission. All data needed to evaluate the conclusions in the paper are openly available in the paper and the Supplementary Materials, and in the ``Source code and data for AIMMD and PE estimate'' repository at \url{http://doi.org/10.5281/zenodo.8048453}.

\section*{Code Availability Statement}

We performed path sampling simulations adapting the AIMMD Python package developed by Jung\cite{jung2023machine}, which builds upon OpenPathSampling (OPS), a Python library  for TPS simulations\cite{swenson2018openpathsampling1}.
We performed the reweighting and projections described in  Section~\ref{sec:theoryE} with the custom-written PathEnsemble Python package. The PathEnsemble code, the data featured in this paper, and the scripts for running the simulations and analyzing the results are available at the repository \href{http://doi.org/10.5281/zenodo.8048453}{DOI:10.5281/zenodo.8048453}.

\appendix

\section{\label{app1}Crossing probability along the committor}

We demonstrate Eq.~\eqref{crossing0} in the non-restrictive assumption of Markovian dynamics\cite{best2011diffusion,berezhkovskii2011time}. If $\lambda=\lambda_{\mathrm{A}}$ or $\lambda=1$, the proof is trivial---the last one follows from the definition of committor.
%(?)
Assume now that $0<\lambda_{\mathrm{A}}<\lambda<1$; $\textbf x$ is a  trajectory leaving $\mathrm{A}$ at $t=0$ and crossing the committor value $\lambda_{\mathrm{A}}$. Let $t'$ be the first time when $p_{\mathrm{B}}(\textbf x(t'))=\lambda_{\mathrm{A}}$. $\textbf{x}$ can continue from $t'$ in 3 possible ways:
\begin{itemize}
    \item[1.] it reaches $\mathrm{A}$ before $\mathrm{B}$ without crossing $\lambda$;
    \item[2.] it reaches $\mathrm{A}$ before $\mathrm{B}$ after crossing $\lambda$;
    \item[3.] it reaches $\mathrm{B}$ before $\mathrm{A}$, hence crossing $\lambda$,
\end{itemize}
with probability $p_1$, $p_2$, and $p_3$, respectively. We find that $p_1=1-P_{\mathrm{A}}(\lambda\mid \lambda_{\mathrm{A}})$, $p_3 = p_{\mathrm{B}}(\textbf x(t'))=\lambda_{\mathrm{A}}$, and $p_2=P_{\mathrm{A}}(\lambda\mid \lambda_{\mathrm{A}})~(1-\lambda)$ as combination of two independent events: $\textbf x$ crossing $\lambda$ from $\lambda_{\mathrm{A}}$ before $\mathrm{A}$ and $\textbf x$ reaching $\mathrm{A}$ from $\lambda$ before $\mathrm{B}$.
% definition of Markovianity
Since $p_1+p_2+p_3=1$:
\begin{equation}
\begin{aligned}
    1 - P_{\mathrm{A}}(\lambda\mid \lambda_{\mathrm{A}}) + \lambda_{\mathrm{A}} + P_{\mathrm{A}}(\lambda\mid \lambda_{\mathrm{A}})~(1-\lambda) = 1\\
    P_{\mathrm{A}}(\lambda\mid \lambda_{\mathrm{A}})~(- 1 + 1 - \lambda) + \lambda_{\mathrm{A}} = 0
\end{aligned}
\end{equation}
from which $P_{\mathrm{A}}(\lambda\mid \lambda_{\mathrm{A}})=\lambda_{\mathrm{A}}/\lambda$.
A related results was obtained in Eq.~9 of Ref.~\onlinecite{vanden2008assumptions}.

\section{\label{app3}Connection to the RPE theory}

We explain the connection between the weighting scheme described in Eq.~\eqref{weightsA} with TIS\cite{van2005elaborating} and the RPE approach\cite{rogal2010reweighted}. In previous studies, the transition region between $\mathrm{A}$ and $\mathrm{B}$ was partitioned into interfaces defined by a progress coordinate, which generally was not the committor. TIS then required sampling a large number of unbiased trajectories at each interface. From these simulations, one can estimate an ensemble for each interface and then merge them according to the global crossing probability estimated, e.g., with WHAM\cite{kumar1992weighted,stelzl_dynamic_2017,ferguson_bayeswham_2017}).
This method proved to be computationally demanding and highly sensitive to the interface selection\cite{kratzer2013automatic}.

Recently, Brotzakis and Bolhuis developed the virtual interface exchange (VIE) algorithm, which populates the TIS interfaces with TPS trial trajectories\cite{brotzakis2019approximating,coluzza2005virtual,frenkel2006wasterecycling}. Having access to the committor, the optimal reaction coordinate\cite{berezhkovskii2022relations}, our method can be seen as a limit case of VIE with an infinite number of interfaces defined after the committor.

Carrying the analogy with the RPE approach, each AIMMD trial trajectory $\textbf x^{(i)}$ is now the unique representative of the interface
\begin{equation}
    I_{\lambda^{(i)}}=\{ x \mid p_{\mathrm{B}}(x) = \lambda^{(i)}\},
\end{equation}
defined by its shooting point, with $p_{\mathrm{B}}(x_{\mathrm{sp}}^{(i)})=\lambda^{(i)}$. Once the AIMMD sampling has converged, $x_{\mathrm{sp}}^{(i)}$ is proportional to the Boltzmann distribution restricted to $I_{\lambda^{(i)}}$, since it is drawn from the TPE distribution:
\begin{equation}
\begin{aligned}
    \rho_{\mathrm{TPE}}(x)&\propto \rho(x)~P(\mathrm{TP}\mid x) \\&\propto \rho(x)~p_{\mathrm{B}}(x)~(1-p_{\mathrm{B}}(x)),
\end{aligned}
\end{equation}
and the target selection probability does not alter $\rho_{\mathrm{TPE}}$ within $I_{\lambda^{(i)}}$. We retain the initial TPS steps (before convergence) in the computations to optimize resource usage.

The RPE combines all the associated interfaces by assigning weights related to the crossing probability\cite{rogal2010reweighted}. Each interface $I_{\lambda^{(i)}}$ gives its own ``crossing statistics'' (or histogram) starting from $\lambda^{(i)}$:
\begin{equation}
H_{\mathrm{A}}^{(i)}(\lambda) = \tilde h_{\mathrm{A}}[\textbf x^{(i)}]~\theta(\lambda-\lambda^{(i)})~\theta(\lambda_{\mathrm{max}}^{(i)}-\lambda),
\end{equation}
in which $\textbf x^{(i)}$ is the only contributor. The statistics before $\lambda^{(i)}$ do not provide meaningful information because we deliberately forced the trajectory to reach that point.

The total crossing statistics from $\lambda_{\mathrm{A}}$ can be obtained in two ways:
\begin{itemize}
    \item[1.] from the $H_{\mathrm{A}}^{(i)}(\lambda)$, through a function $f(\lambda)$:
    \begin{equation}\label{combined_histograms}
        H_{\mathrm{A}}(\lambda) = f(\lambda)\sum_{i=1}^{n_{\mathrm{TPS}}} H_{\mathrm{A}}^{(i)}(\lambda);
    \end{equation}
    \item[2.] from the individually weighted trajectories (as in Fig.~\ref{fig:2}d):
    \begin{equation}\label{combined_trajectories}
    K_{\mathrm{A}}(\lambda) = \sum_{i=1}^{n_{\mathrm{TPS}}} w_{\mathrm{A}}^{(i)}~\theta(\lambda_{\mathrm{max}}^{(i)}-\lambda).
    \end{equation}
\end{itemize}
The two approaches are apparently very different: $H^{(i)}_{\mathrm{A}}$ does not contribute to the total statistics before its shooting interface, as $H_{\mathrm{A}}^{(i)}(\lambda<\lambda^{(i)})=0$, whereas $\textbf x^{(i)}$ does.

Rogal et al.\cite{rogal2010reweighted,brotzakis2019approximating} showed that $K_{\mathrm{A}}(\lambda)$ converges to $H_{\mathrm{A}}(\lambda)$ if we set
\begin{equation}\label{correct_weights}
    w_{\mathrm{A}}^{(i)} = \tilde h_{\mathrm{A}}[\textbf x^{(i)}]~f(\lambda_{\mathrm{max}}^{(i)})
\end{equation}
in the limit of infinite trajectories. Therefore, Eq.~\eqref{correct_weights} gives the optimal weights for reconstructing $\mathcal P_{\mathrm{A},\lambda_{\mathrm{A}}}^+$. Here, we impose $H_{\mathrm{A}}(\lambda)$ to match $P_{\mathrm{A}}(\lambda\mid \lambda_{\mathrm{A}})$. By comparing Eqs.~\eqref{crossing0} and \eqref{combined_histograms}, we obtain
\begin{equation}
    f(\lambda) \propto \frac{1}{\lambda~m_{\mathrm{A}}(\lambda)},
\end{equation}
where $m_{\mathrm{A}}$ is defined as in Eq.~\eqref{mA}. By injecting Eq.~\eqref{correct_weights}, we finally get Eq.~\eqref{weightsA}.

\section{\label{app2}Normalization constants of the RPE}

We derive the constraint of Eq.~\eqref{normalization}. Bolhuis and Lechner\cite{bolhuis2011relation} proved that the effective committor function of the coordinates $q$ is
\begin{equation}\label{effective_committor}
    p_{\mathrm{B}}(q) = \frac{\rho_{\mathrm{B}}(q)}{\rho_{\mathrm{A}}(q)+\rho_{\mathrm{B}}(q)}. 
\end{equation}
If the $q=\lambda$ are committor values themselves, then $p_{\mathrm{B}}(\lambda)=\lambda$. At the transition state ($\lambda=0.5$) we obtain the identity: 
\begin{equation}\label{identity}
    \frac{1}{2} = \frac{\rho_{\mathrm{B}}(\lambda=0.5)}{\rho_{\mathrm{A}}(0.5)+\rho_{\mathrm{B}}(0.5)},
\end{equation}
satisfied by $\rho_{\mathrm{A}}(0.5)=\rho_{\mathrm{B}}(0.5)$ and $c_{\mathrm{A}}~\tilde \rho_{\mathrm{A}}(0.5)=c_{\mathrm{B}}~\tilde \rho_{\mathrm{B}}(0.5)$.

%\nocite{*}
\bibliography{main}

\newpage
\begin{figure*}
\renewcommand{\thefigure}{S1}
    \centering
    \includegraphics[width=.34\textwidth]{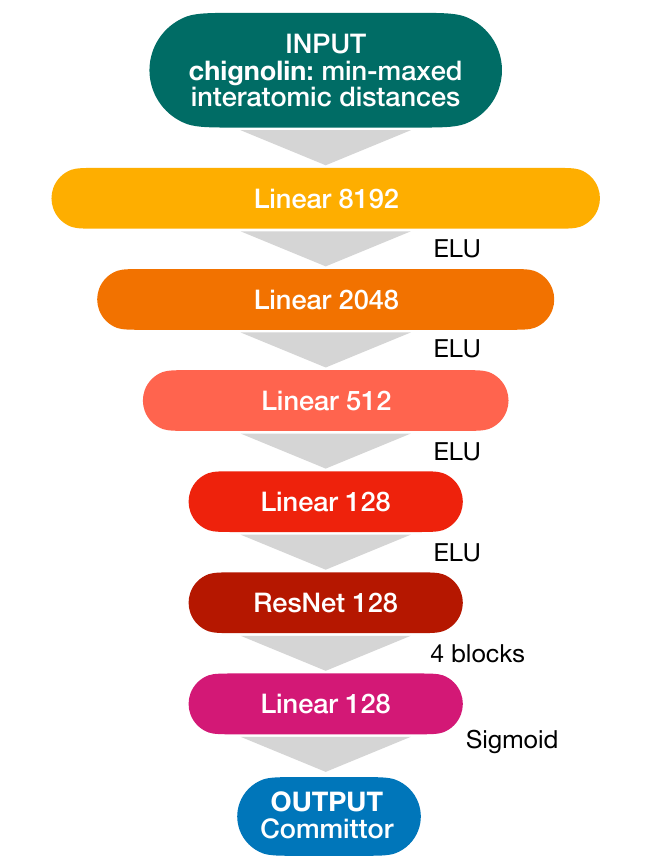}
    \caption{Neural network architecture. The first block is valid only for the application on chignolin. For the 2D system, we directly feed the $x,y$ coordinates to the neural network and leave the rest unchanged.}
    \label{fig:S1}
\end{figure*}

\begin{figure*}
\renewcommand{\thefigure}{S2}
    \centering
    \includegraphics[width=\textwidth]{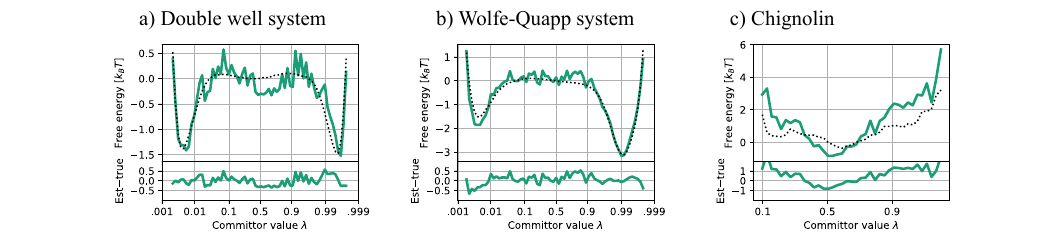}
    \caption{Free energy of the TPE obtained by AIMMD run1 for the double well (\textbf{a}), Wolfe-Quapp (\textbf{b}), and chignolin (\textbf{c}). We used 500 (\textbf{a}, \textbf{b}) and 50 steps (\textbf{c}). The dotted lines are the reference profiles from numerical computations (2D systems) and long equilibrium MD simulations (chignolin).}
    \label{fig:S2}
\end{figure*}

\begin{figure*}
\renewcommand{\thefigure}{S3}
    \centering
    \includegraphics[width=\columnwidth]{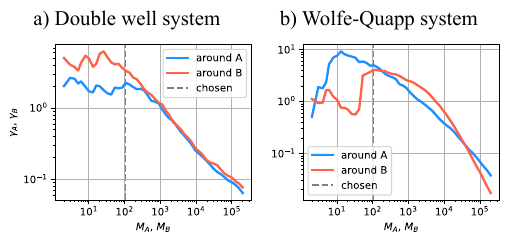}
    \caption{Determining the weighting factors for the equilibrium simulations in the states, for the double well (\textbf{a}) and Wolfe-Quapp system (\textbf{b}). The plot shows the weighting factors $\gamma_{\mathrm{A}}$ (solid blue) and $\gamma_{\mathrm{B}}$ (solid red) as a function of the $M_{\mathrm{A}},~M_{\mathrm{B}}$ determining the $\lambda_{\mathrm{A}},~\lambda_{\mathrm{B}}$ thresholds. The dashed lines mark the chosen $M_{\mathrm{A}}=100$ and $M_{\mathrm{B}}=100$, leading to more robust factors.}
    \label{fig:S3}
\end{figure*}

\begin{figure*}
\renewcommand{\thefigure}{S4}
    \centering
    \includegraphics[width=\textwidth]{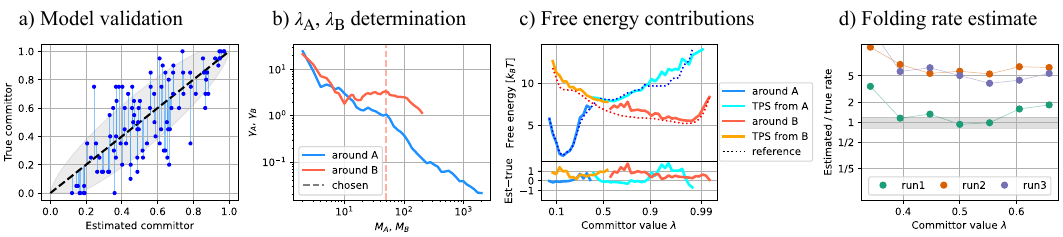}
    \caption{Chignolin, AIMMD run1, first 50 steps. \textbf{a}) Committor model tested on an independent shooting points validation set. The ``true committor'' is the outcome of 20 independent shots from each point, with its 95\% confidence interval denoted by the gray region. \textbf{b}) Weighting factors $\gamma_{\mathrm{A}}$ (solid blue) and $\gamma_{\mathrm{B}}$ (solid red) as a function of the $M_{\mathrm{A}},~M_{\mathrm{B}}$ determining the $\lambda_{\mathrm{A}},~\lambda_{\mathrm{B}}$ thresholds. The dashed lines mark the chosen $M_{\mathrm{A}}=10$ (blue) and $M_{\mathrm{B}}=50$ (red). \textbf{c}) Free energy profiles of the 4 contributions to the density in Eq.~\eqref{densities}. The dotted lines are the reference $F_{\mathrm{A}}(q)=-k_{\mathrm{B}}T~\log\rho_{\mathrm{A}}$ (blue) and $F_{\mathrm{B}}(q)=-k_{\mathrm{B}}T~\log\rho_{\mathrm{B}}$ (red) from the equilibrium MD simulations. The free energy offsets are determined by the weighting factors $\gamma_{\mathrm{A}}$ and $\gamma_{\mathrm{B}}$. \textbf{d}) Bayesian folding rate estimate ($k_{\mathrm{BA}}=k_{\textit{fold}}$) at different committor values. The gray area is the 95\% confidence interval of $k_{\mathrm{BA}}$ from long equilibrium MD simulations.}
    \label{fig:S4}
\end{figure*}

\begin{figure*}
\renewcommand{\thefigure}{S5}
    \centering
    \includegraphics[width=\columnwidth]{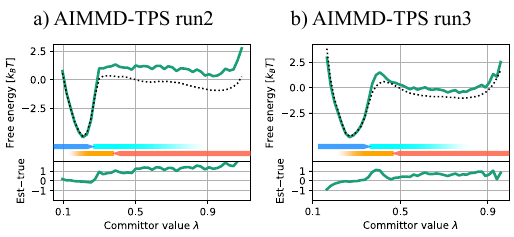}
    \caption{Chignolin's folding free energy profiles for runs 2 (\textbf{a}) and 3 (\textbf{b}) of AIMMD. The plots show the free energy as a function of the committor  estimated after 50 steps (solid line, top axis), free energy from the equilibrium simulations (dotted line), and difference between the two (bottom axis).}
    \label{fig:S5}
\end{figure*}

\begin{figure*}
\renewcommand{\thefigure}{S6}
    \centering
    \includegraphics[width=\textwidth]{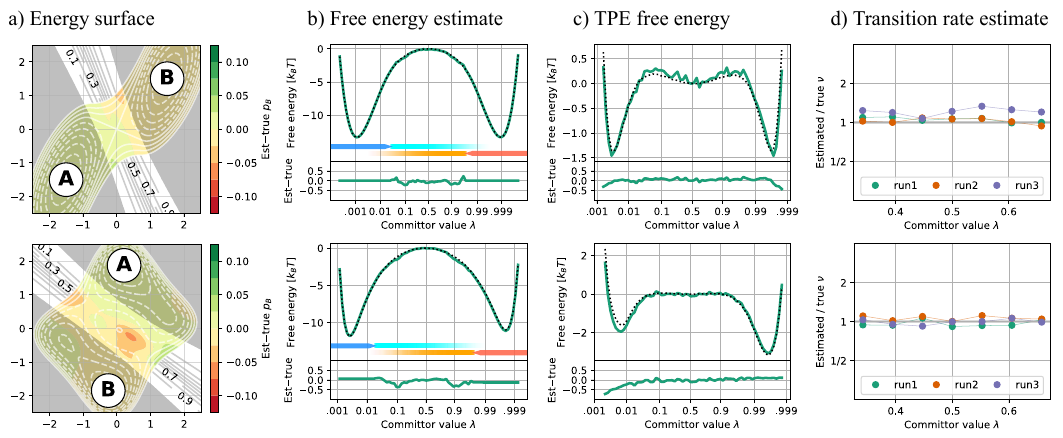}
    \caption{2D systems in the data-rich regime (top: double well, bottom: Wolfe-Quapp). Results obtained after 5,000 AIMMD steps. \textbf{a)} Run1, committor model (contour lines), error of the model (filled contour), and region between the $\lambda_{\mathrm{A}}$ and $\lambda_{\mathrm{B}}$ cutoffs (light area). \textbf{b)} Run1, estimated free energy (solid line, top axis), reference free energy from numerical computation (dotted line), and error of the estimate (bottom axis). The arrows indicate the contributions of the simulations around A (blue), the TPS trajectories from A (cyan), the simulations around B (red), and the TPS trajectories from B (orange). \textbf{c)} Run1, TPE free energy (solid line, top axis), numerical profile (dotted line), and error of the estimate (bottom axis). \textbf{d)} Bayesian rate estimate of $\nu$ at different committor values, each color denoting a different run. The gray area is the 95\% confidence interval of $\nu$ from the equilibrium simulations.}
    \label{fig:S6}
\end{figure*}

\begin{figure*}
\renewcommand{\thefigure}{S7}
    \centering
    \includegraphics[width=\textwidth]{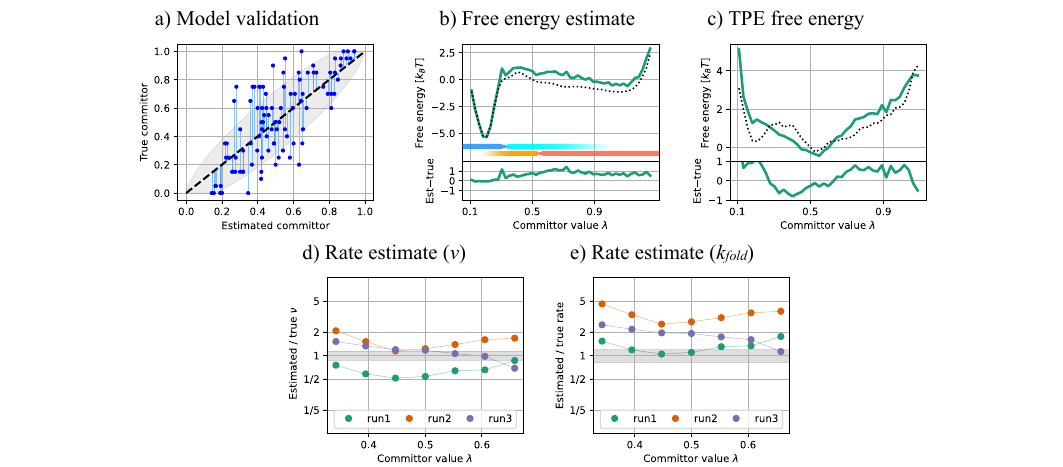}
    \caption{Chignolin, results after 250 AIMMD steps. \textbf{a)} Run1, estimated committor tested on an independent shooting points validation set. The ``true committor'' is the outcome of 20 independent shots from each point, with its 95\% confidence interval denoted by the gray region. \textbf{b)} Run1, estimated free energy (solid line, top axis), reference free energy from numerical computation (dotted line), and error of the estimate (bottom axis). The arrows indicate the contributions of the simulations around A (blue), the TPS trajectories from A (cyan), the simulations around B (red), and the TPS trajectories from B (orange). \textbf{c)} Run1, estimated TPE free energy (solid line, top axis), reference numerical profile (dotted line), and error of the estimate (bottom axis). \textbf{d)} Bayesian rate estimate of $\nu$ at different committor values, each color denoting a different run. The gray area is the 95\% confidence interval of $\nu$ from long equilibrium simulations. \textbf{e)} Bayesian folding rate estimate ($k_{\mathrm{BA}}=k_{\textit{fold}}$), represented as in \textbf{d}.}
    \label{fig:S7}
\end{figure*}

\begin{figure*}
\renewcommand{\thefigure}{S8}
    \centering
    \includegraphics[width=\columnwidth]{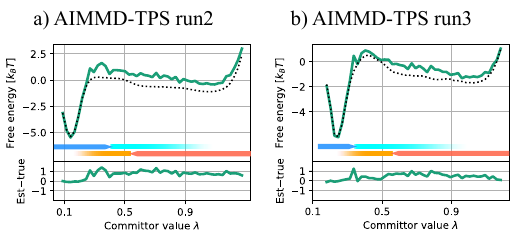}
    \caption{Chignolin's folding free energy profiles for runs 2 (\textbf{a}) and 3 (\textbf{b}) of AIMMD. The plots show the free energy as a function of the committor  estimated after 250 steps (solid line, top axis), free energy from the equilibrium simulations (dotted line), and difference between the two (bottom axis).}
    \label{fig:S8}
\end{figure*}

\begin{figure*}
    \renewcommand{\thefigure}{S9}
    \centering
    \includegraphics[width=\textwidth]{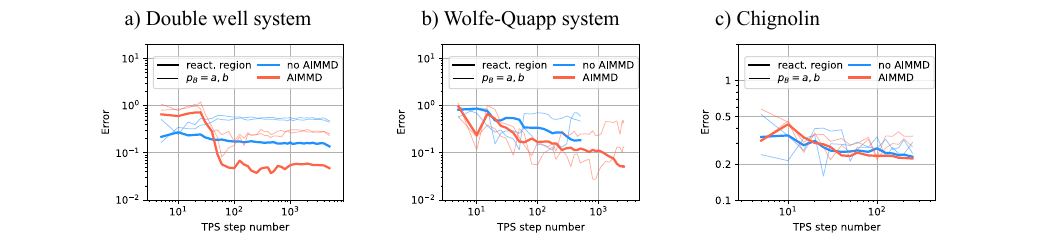}
    \caption{Error on the estimated committor as a function of an increasing number of steps for the double well (\textbf{a}), Wolfe-Quapp (\textbf{b}) and chignolin (\textbf{c}). The red lines track the evolution of the model from AIMMD run1, while the blue lines correspond to neural networks trained \textit{a posteriori} on the TPS run0 data. In the double well and Wolfe-Quapp, the thick and narrow lines denote the error in the transition region, and at the $\lambda_{\mathrm{A}}$ and $\lambda_{\mathrm{B}}$ interfaces, respectively. For each model, we took a validation set of TPE configurations uniformly distributed in the reference $p_{\mathrm{B}}$ space; we then computed the RSME of $q(x)=\mathrm{logit}~p_{\mathrm{B}}(x)$, and rescaled the result by 4. In this way, small $q(x)$ errors at the TS are approximately the absolute error of $p_{\mathrm{B}}(x)$; furthermore, $p_{\mathrm{B}}(x)$ errors close to the states are penalized as they have a larger impact on the free energy accuracy.}
    \label{fig:S9}
\end{figure*}

\begin{figure*}
    \renewcommand{\thefigure}{S10}
    \centering
    \includegraphics[width=\textwidth]{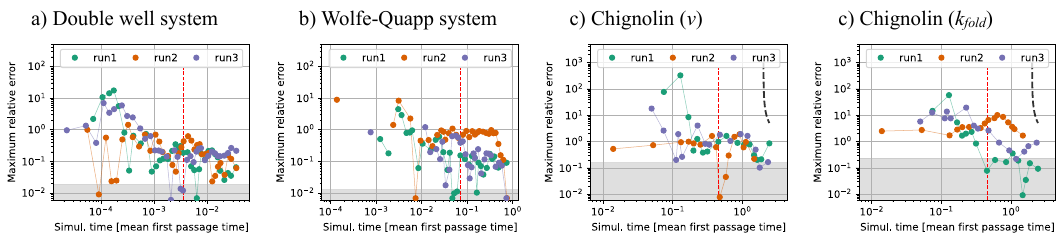}
    \caption{Accuracy of rate estimates as a function of simulated time for the double well (\textbf{a}), Wolfe-Quapp (\textbf{b}) and chignolin (\textbf{c}). The plots show the maximum relative error of the $\nu$ estimate at $\lambda=0.5$ (transition state) with the total simulated time in units of the system's mean first passage time. Each color corresponds to a different run. For chignolin, we also plot the error of the folding rate $k_{\mathrm{BA}}=k_{\textit{fold}}$ (\textbf{d}). The vertical lines mark the time of run1 in Figs.~\ref{fig:3},~\ref{fig:5}. The gray area is the 95\% confidence interval of the rates from long equilibrium simulations.
    The dashed line is the expected error when estimating the rates from an equilibrium simulation of corresponding duration.
    }
    \label{fig:S10}
\end{figure*}

\begin{figure*}
    \renewcommand{\thefigure}{S11}
    \centering
    \includegraphics[width=\textwidth]{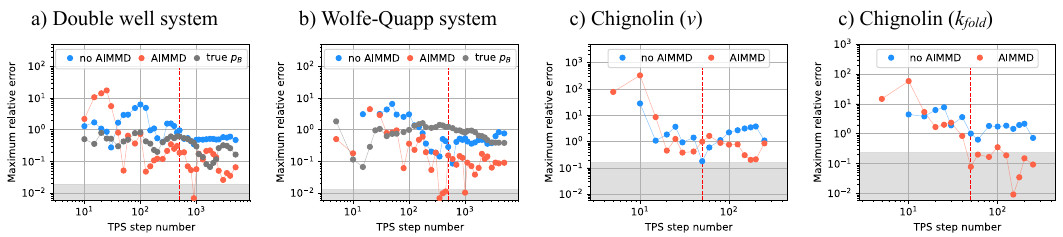}
    \caption{Improvement in the estimation of the rates given by AIMMD, for the double well (\textbf{a}), Wolfe-Quapp (\textbf{b}) and chignolin (\textbf{c}). The plots show the maximum relative error of the $\nu$ estimate at $\lambda=0.5$ (transition state) with the number of TPS steps. Red: original results from AIMMD run1; gray: replacing the neural network with the reference numerical solution of the committor; blue: standard TPS run with a neural network trained \textit{a posteriori}. For chignolin, we also plot the error on the folding rate $k_{\mathrm{BA}}=k_{\textit{fold}}$ (\textbf{d}). The vertical lines mark the step numbers considered in Figs.~\ref{fig:3}~and~\ref{fig:5}. The gray area is the 95\% confidence interval of the rates from long equilibrium simulations.}
    \label{fig:S11}
\end{figure*}

\end{document}